\documentclass[cmpm,envcountsect,envcountsame,numbook]{svjourm}

\usepackage{graphics}

\usepackage{latexsym, amssymb, enumerate, amsmath}

  \usepackage{color}
\usepackage[latin1]{inputenc}
\usepackage[english]{babel}

\usepackage{mathrsfs}
\usepackage[T1]{fontenc}

\usepackage{epsfig}
\usepackage{graphicx}
\usepackage{bm}

\smartqed

\newcommand{\N}{\mathbb{N}}
\newcommand{\R}{\mathbb{R}}
\newcommand{\C}{\mathbb{C}}
\newcommand{\Z}{\mathbb{Z}}
\newcommand{\T}{\mathbb{T}}

\begin{document}
\title{Torus as phase space: Weyl quantization, dequantization and Wigner formalism}
\author{Marilena Ligab\`o % etc
\thanks{\emph{Present address:} marilena.ligabo@poliba.it}%
}                     % Do not remove
\institute{Dipartimento di Meccanica, Matematica e Management -
 Politecnico di Bari 
 \newline
 Via Orabona, 4 - 70125 Bari}
\date{}
% Add name of the expert who has communicated your paper
\communicated{name}
\maketitle
\begin{abstract}
The Weyl quantization of classical observables on the torus (as phase space) without regularity assumptions is explicitly computed.  The equivalence class of symbols yielding the same Weyl operator is characterized.   The Heisenberg equation  for the dynamics of general quantum observables is written through the Moyal brackets on the torus and the support of the Wigner transform is  characterized.  Finally,  a dequantization procedure is introduced that applies, for instance, to the Pauli matrices. As a result we obtain the corresponding classical symbols.
\end{abstract}

\section{Introduction}\label{intro}

In this paper we consider the quantization of systems having the torus $\mathbb{T}^2=\mathbb{R}^2 / \mathbb{Z}^2$ as  classical phase space.   This subject was introduced by Berry and Hannay in \cite{berry} and since then has received much attention from the mathematical community as well as from the physical  one. The mathematical literature (see e.g.  \cite{de,DEGI,Keating1,Keating2,Bou,Bonechi2,Bonechi1,Faure2,Faure1}) mostly deals with  quantization of  {\it (linear) hyperbolic symplectomorphisms}, with the aim of understanding the  quantum counterpart of the classical chaotic behavior.   On a more physical point of view, the discrete  Wigner transform and the finite dimensional Weyl systems have been studied by several authors, (see e.g. \cite{Schwinger,Cohen,Wootters3,Cohendet,Galetti,var,Leonhardt2,Paz1,manko,Wootters2,Paz2,Wootters1,erc,marmo,Zak}) with particular attention  to the implementation of  tomographic techniques on finite dimensional quantum systems. 
\vskip 4pt\noindent
 Referring the reader e.g.  to \cite{deb1,bookdegr,deb2} for  reviews on  the (equivalent) quantization procedures, to be briefly described  in the Appendix, we limit ourselves here to recall that for $h=1/N$ there are infinitely many inequivalent $N$-dimensional unitary representations of the discrete Heisenberg group, labeled by a parameter $\theta \in \T^2$.   These representations yield the quantization of sufficiently regular classical observables via Fourier expansion, and of the linear symplectomorphisms via commutativity between quantization and linear evolution (exact Egorov theorem).  
  
  We note, however, that several relevant mathematical questions concerning the quantization procedure are still open. Namely:
  \begin{itemize}
  \item [(i)] The identification  of the most general  class of classical observables on the torus which can be quantized;
  \item [(ii)] The characterization of the equivalence class of symbols with the same Weyl operator;
 \item[(iii)]
 The determination  of the support of the Wigner transform;
  \item
 [(iv)]
 The quantization of {\it all} sufficiently regular classical dynamical systems (discrete and  continuous in time, linear and nonlinear) on $\T^2$;
 \item
 [(v)]
The dequantization of the quantum observables, i.e. the identification of the classical symbol (function defined on $\T^2$) whose canonical quantization reproduces the given  quantum operator.
\end{itemize}
The purpose of this article is to answer  the above questions.  More specifically:
\begin{itemize}
\item
[(i)] 
(Section 2)
\newline
As in  \cite{DEGI}  the Weyl quantization of a sufficiently smooth function $\alpha$ on $\T^2$ (classical observable) is defined by replacing exponentials in  the Fourier series of $\alpha$ by their representations in $\C^N$,  and  depends on the parameter $\theta=(\theta_1, \theta_2) \in \T^2$ labeling  the chosen representation.  The  corresponding Weyl operator $A$ ($N \times N$ matrix) is explicitly computed, and  depends only on the values of $\alpha$  on the lattice $L(\theta,N):=\left\{\left(\frac{j}{2N}+\frac{\theta_1}{ N}, \frac{k}{2N}+\frac{\theta_2}{ N} \right): j,k \in \Z_{2N}\right\}$ (Theorem \ref{th:weylquant}). By this property all functions $\alpha :\T^2\to\C$ admitting Weyl quantization with respect to the selected representation can be characterized:  $\alpha$ admits quantization if and only if it takes  finite values on $L(\theta,N)$. The quantization map is not  one-to-one:  two functions assuming the same values on $L(\theta,N)$ yield  the  same quantized operator (Corollary \ref{corollarynotinjective}). A less obvious remark is that two functions on $\T^2$ assuming   different  values on $L(\theta,N)$ can generate the same quantum operator. A general procedure is exhibited (Theorem \ref{th:characteq}) to construct functions  assuming different values on the lattice  but generating the same operator upon quantization. 

Within the Weyl quantization, the Moyal product and the Moyal brackets of two classical functions on $\T^2$ are defined (Definitions \ref{defn: Moyal pr} - \ref{defn: Moyal br}) and explicitly computed (Theorems \ref{th:moyal product} - \ref{Th:Moyal br}). This allows us to write  the equation of motion on the classical phase space corresponding to the Heisenberg equation for the evolution of quantum observables on  $\C^N$ (Subsection \ref{Quantum dynamics in phase space: evolution of symbols on the torus}). In this sense any classical dynamics on the torus can be quantized, not only the linear maps that describe stroboscopic, discrete-time dynamics. 
\\

\item[(ii)]
(Section 3)
\newline
The construction of the Wigner transform $W_{\theta,N}\psi$  of a quantum state $\psi \in \C^N$  on $\T^2$ is presented here starting from any chosen representation of the discrete Heisenberg group following the approach in \cite{DEGI}, with some suitable modifications. The resulting $W_{\theta,N}\psi$ is shown to be a  distribution (signed measure) on phase space with total mass equal to one and support $L(\theta,N)$, (Theorem \ref{th:explicitWigner}); moreover  the corresponding  marginals are the position and momentum probability distributions with support $\mathscr{L}_1(\theta_1,N):=\{\frac{j}{N}+\frac{\theta_1}{ N}: j \in \Z_{N}\}$ and $\mathscr{L}_2(\theta_2,N):=\{\frac{j}{N}+\frac{\theta_2}{ N}: j \in \Z_{N}\}$ respectively, (Proposition \ref{propT}).  An equivocation on the support of the Wigner transform $W_{\theta,N}\psi$ is present in the literature  because  the vector $\psi \in \C^N$ is characterized by $N$ complex numbers, and, heuristically,  the  Wigner transform,  quadratic in $\psi$,  is characterized by $N^2$  distinct values and so its support is naively expected to contain $N^2$ points.
In this article this ambiguity is definitely clarified: the correct support of $W_{\theta,N}\psi$ is $L(\theta,N)$, consisting of $4N^2$ points, that strictly contains $\mathscr{L}_1(\theta_1,N) \times \mathscr{L}_2(\theta_2,N)$ (cartesian product of the support of the two marginals). However, the independent values of $W_{\theta,N}\psi$ are only $N^2$, thus, in principle, it can be always restricted to a proper $N \times N$ points lattice, but this restriction is not natural because  the corresponding marginals  are no longer the position and momentum probability measure,  and  the total mass is no longer one.  The larger support was first noticed by Berry and Hannay, \cite{berry},  and recently the doubling feature of the lattice has been reconsidered in the physical  literature, \cite{Paz1,Zak}, and advocated by different physical motivations, through a construction of the discrete Wigner function not based on the representations of the discrete Heisenberg group. 
\\

\item[(iv)]
(Sections 4 and 5)
\newline
 A dequantization method based on the Wigner formalism is introduced: given any  operator $A$ on the Hilbert space $\C^N$,  a function $\alpha: \T^2\to \C$ is identified such that its Weyl quantization is exactly $A$ (Theorem \ref{th:deq}). An example is the dequantization of the Pauli matrices  $\sigma_x, \sigma_y, \sigma_z$ in the case of a spin $1/2$, i.e. $N=2$.   Moreover,  the values of the Wigner transform of a generic state $\psi \in \C^2$ are shown to be  the expectation values of the operators $\sigma_x, \sigma_y, \sigma_z$ in the state $\psi$, with appropriate symmetries. This correspondence  is  generalized to any  $N$, and, as a byproduct, this procedure yields, in a natural way, $N$-dimensional versions of the Pauli matrices (Subsection \ref{Npauli}).
\end{itemize}

\section{Weyl quantization on the torus}
The standard Weyl quantization on the torus consists in a map that associates to a smooth function defined on the torus $\T^2=\R^2 / \Z^2$ a quantum operator on the Hilbert space $ \C^N$, where $N$ is related to the Planck constant $h$ via the relation $h=\tfrac{1}{N}$. 

The Weyl quantization is obtained selecting one $N$-dimensional representation of the discrete Heisenberg group $\mathbb{H}(\Z)$ from the family $\{T_{\theta,N}\}_{\theta \in \T^2}$, where for all $\theta=(\theta_1,\theta_2) \in \T^2$ the unitary representation $T_{\theta,N}: \mathbb{H}(\Z) \to \mathcal{U}(\C^N)$ acts as follows on the canonical basis $\{u_j\}_{j=0, \dots,N-1}$ of $\C^N$
\begin{equation}\label{defrepresteta}
T_{\theta,N}(n_1,n_2,s)u_j=e^{\frac{2\pi i s}{N}}e^{-\frac{\pi in_1n_2}{N}}e^{\frac{2\pi i n_1 (j+\theta_1)}{N}}e^{\frac{2\pi i n_2 \theta_2}{N}}u_{j-n_2},
\end{equation}
for all $n_1,n_2 \in \Z$ and $s \in \R$, where the result of $j-n_2$ is  modulo $N$, and $\mathcal{U}(\C^N)$ denotes the space of unitary operators on $\C^N$. 
The variable $s$ acts just as a scalar multiplication by $e^{\frac{2\pi i s}{N}}$, hence it  can be disregarded and, with an abuse of notation, we can define the map $T_{\theta,N}:\Z^2 \to \mathcal{U}(\C^N)$:  
\begin{equation}
T_{\theta,N}(n_1,n_2):=T_{\theta,N}(n_1,n_2,0), \qquad n_1,n_2 \in \Z.
\end{equation}
For a short review of the representation theory of the discrete Heisenberg group $\mathbb{H}(\Z)$ we refer to the Appendix and to the references therein. In what follows we select one representation $T_{\theta,N}$ by choosing $N \in \N$, $N \geq 1$, and $\theta = (\theta_1, \theta_2)\in \T^2$ and we refer to $T_{\theta,N}$ as \emph{representation $(\theta,N)$}.
\par\noindent
Let $\alpha\in C^1(\T^2;\C)$. By the Fourier inversion formula:
\begin{equation}\label{fourierinv}
\alpha(x,p)=\sum_{n_1, n_2\in \Z}\hat{\alpha}(n_1, n_2)e^{2\pi i (n_1 x+n_2p)}, \qquad (x,p) \in \T^2,
\end{equation}
where
\begin{equation}
\hat{\alpha}(n_1, n_2)=\int_{\T^2} \alpha(x,p)e^{-2\pi i (xn_1+ p n_2)}\; \mathrm{d}x \; \mathrm{d}p , \qquad (n_1,n_2) \in \Z^2. \nonumber
\end{equation}
and the Fourier series in (\ref{fourierinv}) converges uniformly.
\begin{definition}\label{def:Weyl op}
The Weyl quantization  of $\alpha \in C^1(\T^2;\C)$ in the representation $(\theta,N)$, denoted by  $\textrm{Op}_{\theta,N}^W(\alpha)$, is defined as  follows
\begin{equation}
\textrm{Op}_{\theta,N}^W(\alpha)=\sum_{n_1, n_2\in \Z}\hat{\alpha}(n_1, n_2)T_{\theta,N}(n_1,n_2). \nonumber
\end{equation}
\end{definition}
\begin{remark}
By Definition \ref{def:Weyl op} it follows immediately that for all $n_1,n_2 \in \Z$:
\begin{equation}\label{symbolsof generators}
T_{\theta,N}(n_1,n_2)=\textrm{Op}_{\theta,N}^W(e^{2\pi i (n_1x +n_2p)}).\nonumber
\end{equation}
\end{remark}

The first result is  the explicit computation of the  matrix elements of the Weyl operators. 

We introduce some notations: let $\{u_j\}_{j=0,\ldots,N-1}$ be the canonical basis in $\C^N$, it is convenient to regard the index $j$ as   an element of $\Z_N$, with $\Z_{N}= \Z/ N \Z$. Let $\langle \cdot, \cdot \rangle$ denote the scalar product on $\C^N$, and $M_N(\C)$ the space of the $N \times N$ complex matrices. If $A \in M_N(\C)$ we denote with $A_{j,k}$ its entries, $j,k=0, \dots, N-1$. It is convenient to regard the indices $j$ and $k$ as elements of $\Z_N$. Finally we denote with  $\mathcal{F}_2$  the discrete Fourier transform with respect to the second variable, namely:  if $\beta: \Z_{2N} \times \Z_{2N} \to \C$ then
\begin{equation}
\mathcal{F}_2 \beta(m, r)=\sum_{j\in \Z_{2N}}\beta(m,j)\;e^{-\frac{2\pi i r j}{2N}}, \quad m,r \in \Z_{2N}.
\end{equation}
\begin{theorem}\label{th:weylquant}
Let $\alpha\in C^1(\T^2; \C)$. Then for all $n,j = 0,\dots ,N-1$
\begin{eqnarray}\label{matrixcoeffWeylquant}
&& \langle u_n, \textrm{Op}_{\theta,N}^W(\alpha)u_j \rangle \nonumber \\
&& =\frac{1}{2N} \left(\mathcal{F}_2 \alpha_{\theta,N }(j+n, j-n)+\mathcal{F}_2 \alpha_{\theta,N }(j+n+N, j-n+N) \right),
\end{eqnarray} 
where for all $m,l \in \Z_{2N}$
\begin{equation}
\alpha_{\theta,N}(m, l):=\alpha\left(  \frac{m}{2N}+\frac{\theta_1}{ N},\frac{j}{2N}+ \frac{\theta_2}{ N}  \right). \nonumber
\end{equation}
\end{theorem}
\proof
We  compute the action of $\textrm{Op}_{\theta,N }^W(\alpha)$ on the vector $u_j$, obtaining:  
\begin{eqnarray}\label{opsigma1}
&& \textrm{Op}_{\theta,N }^W(\alpha)u_j \nonumber \\
&=& \sum_{n_1, n_2\in \Z}\hat{\alpha}(n_1, n_2)T_{\theta,N}(n_1,n_2)u_j \nonumber \\
&=& \sum_{n_1, n_2\in \Z}\hat{\alpha}(n_1, n_2)e^{-\frac{\pi i n_1n_2}{N}}e^{\frac{2\pi i(n_1\theta_1+n_2 \theta_2)}{N}} e^{\frac{2\pi i n_1j}{N}}u_{j-n_2} \nonumber \\
&=&\sum_{n_1, n_2\in \Z} \left(  \int_{\T^2} \alpha(x, p) e^{-2\pi i (n_1x+n_2p)}\; \mathrm{d}x \; \mathrm{d}p  \right) \;\;e^{-\frac{\pi in_1n_2}{N}}e^{\frac{2\pi i(n_1\theta_1+n_2 \theta_2)}{N}} e^{\frac{2\pi i n_1j}{N}} \; u_{j-n_2} \nonumber \\
&=& \mathop{\sum_{m_1, m_2\in \Z}}_{r,s=0,\dots,2N-1}  \int_{\T^2} \alpha(x, p) e^{-2\pi i [(r+2Nm_1)x+(s+2Nm_2)p]} \; \mathrm{d}x \; \mathrm{d}p     \nonumber \\
&& \times e^{-\frac{\pi i(r+2Nm_1)(s+2Nm_2)}{N}}e^{\frac{2\pi i[(r+2Nm_1)\theta_1+(s+2Nm_2) \theta_2]}{N}} e^{\frac{2\pi i (r+2Nm_1)j}{N}} \; u_{j-(s+2Nm_2)} 
\end{eqnarray}
where in (\ref{opsigma1}) we divide $n_1$ and $n_2$ by $2N$, namely we write $n_1=r+2Nm_1$ and $n_2=s+2Nm_2$. Then using the uniform convergence we pass the series inside the integral  obtaining
\begin{eqnarray}\label{eqn:intpass}
&& \textrm{Op}_{\theta,N }^W(\alpha)u_j \nonumber \\
&=&\sum_{r,s=0}^{2N-1}e^{-\frac{\pi i rs}{N}}e^{\frac{2\pi i rj}{N}}  \left( \int_{\T^2} \alpha(x, p) e^{-2\pi i \left[r\left(x-\frac{\theta_1}{ N} \right)+s \left( p -\frac{\theta_2}{ N}\right)\right]} \right.    \nonumber \\
&& \left. \times \sum_{m_1,m_2 \in \Z} e^{-2\pi i \left[ m_1 \left( 2Nx- 2\theta_1\right) +m_2  \left( 2Np-2\theta_2 \right)\right]}\; \mathrm{d}x \; \mathrm{d}p  \right) u_{j-s} .
\end{eqnarray}
Now, using the Poisson summation formula, \cite{Zy}, 
\begin{equation}\label{Poisson Formula}
\sum_{m \in \Z}e^{2\pi i m y}= \sum_{m \in \Z} \delta(y-m),
\end{equation}
for the sum with respect to $m_1,m_2 \in \Z$ in (\ref{eqn:intpass}), we obtain that
\begin{eqnarray}\label{eqnarray:opj}
&& \textrm{Op}_{\theta,N }^W(\alpha)u_j \nonumber \\
&=&\frac{1}{(2N)^2}\sum_{r,s=0}^{2N-1}\left( \int_{\T^2} \alpha(x, p) e^{-2\pi i \left[r\left(x-\frac{\theta_1}{N} \right)+s \left( p -\frac{\theta_2}{ N}\right)\right]} \; \; \mathrm{d}x \; \mathrm{d}p  \right)   e^{-\frac{\pi irs}{N}}e^{\frac{2\pi i rj}{N}}  \nonumber \\
&& \times \sum_{m_1,m_2 \in \Z} \delta \left(x-\frac{1}{2N} \left(m_1+ 2\theta_1\right) \right) \delta \left( p-\frac{1}{2N} \left( m_2+2\theta_2\right)\right) \; u_{j-s} \nonumber \\
&=&\frac{1}{(2N)^2}\sum_{r,s,m_1,m_2=0}^{2N-1}\alpha_{\theta,N}(m_1,m_2)  e^{-\frac{2\pi i(rm_1+sm_2)}{2N} }e^{-\frac{\pi i rs}{N}} e^{\frac{2\pi i rj}{N}} u_{j-s} \nonumber \\
&=&\frac{1}{(2N)^2}\sum_{s,m_1,m_2=0}^{2N-1} \alpha_{\theta,N}(m_1,m_2)  e^{-\frac{2\pi ism_2}{2N} }  \sum_{r=0}^{2N-1}e^{-\frac{2\pi ir(m_1-2j+s)}{2N} } u_{j-s}   \\
&=&\frac{1}{2N}\sum_{s,m_1,m_2=0}^{2N-1} \alpha_{\theta,N}(m_1,m_2)  e^{-\frac{2\pi ism_2}{2N} }\delta^{(2N)}_{s,2j-m_1} u_{j-s}  \nonumber \\
&=&\frac{1}{2N}\sum_{m_1, m_2=0}^{2N-1} \alpha_{\theta,N}(m_1,m_2)  e^{-\frac{2\pi im_2(2j-m_1)}{2N} } u_{m_1-j} \nonumber  ,
\end{eqnarray} 
where in (\ref{eqnarray:opj}) we used the following identity for the sum with respect to $r$:
\begin{equation} 
 \frac{1}{2N}\sum_{j=0}^{2N-1}e^{-\frac{2\pi i j k}{2N} }= \delta^{(2N)}_{k,0},
\end{equation}
where $\delta^{(2N)}_{k,0}$ denotes the Kronecker delta on $\Z_{2N}$.
Therefore
\begin{eqnarray}
&&\langle u_n, \textrm{Op}_{\theta,N }^W(\alpha)u_j \rangle \nonumber \\
 &=& \frac{1}{2N}\sum_{m_1, m_2=0}^{2N-1} \alpha_{\theta,N}(m_1,m_2)  e^{-\frac{2\pi im_2(2j-m_1)}{2N} } \delta^{(N)}_{n,m_1-j} \nonumber \\
&=& \frac{1}{2N}\sum_{m_2=0}^{2N-1}  \left(\alpha_{\theta,N}(j+n,m_2)  e^{-\frac{2\pi im_2(j-n)}{2N} } +  \alpha_{\theta,N}(j+n+N ,m_2)  e^{-\frac{2\pi im_2(j-n+N)}{2N} }\right) \nonumber \\
&=& \frac{1}{2N} \left( \mathcal{F}_2 \alpha_{\theta,N}(j+n,j-n)+  \mathcal{F}_2 \alpha_{\theta,N}(j+n+N,j-n+N)\right) \nonumber
\end{eqnarray}
and this concludes the proof. \qed
\begin{remark}
A simple computation yields
\begin{equation}
\textrm{Op}_{\theta,N}^W(\alpha)^*=\textrm{Op}_{\theta,N}^W(\overline{\alpha}). \nonumber
\end{equation}
Hence the Weyl quantization of any real-valued function is  a self-adjoint operator.
\end{remark}
\begin{remark}
The matrix elements in (\ref{matrixcoeffWeylquant}) are the anaologous of the kernel of the Weyl operator in the well-known Weyl quantization on $\R^2$. If $a: \R^2 \to \C$ is any smooth function then its Weyl quantization is an integral operator on $L^2(\R)$, namely
\begin{equation}
\textrm{Op}^W_{h}(a)\psi(x)= \int_{\R}K_a(x,y) \psi(y)\; \mathrm{d}y, \qquad \psi \in L^2(\R). \nonumber
\end{equation}
The  kernel $K_a$ is
\begin{equation}
K_a(x,y)=\frac{1}{h}\mathscr{F}_2a\left( \frac{x+y}{2}, \frac{y-x}{h}\right)=\frac{1}{h}\int_{\R}a\left(\frac{x+y}{2},p \right)e^{-\frac{2\pi i p(x-y)}{h}}\; \mathrm{d}p , \nonumber
\end{equation}
where $\mathscr{F}_2$ denotes the Fourier transform with respect to the second variable on $\R$.
\end{remark}
\begin{remark}\label{remarknotinjective}
 Theorem \ref{th:weylquant} shows that the Weyl quantization map $\alpha\mapsto Op^W_{\theta,N}(\alpha)$ is not  injective  between ${\mathcal F}(\T^2)$, the space of functions on $\T^2$, and ${\mathcal L}(\C^N) \simeq M_N(\C)$, the space of linear operators acting on $\C^N$. Indeed, any two functions assuming the same values on the lattice 
\begin{equation}\label{defn:latticetheta}
L(\theta,N):=\left\{\left(\frac{j}{2N}+\frac{\theta_1}{ N}, \frac{k}{2N}+\frac{\theta_2}{ N} \right): j,k \in \Z_{2N}\right\} \nonumber
\end{equation} 
yield the same Weyl operator in the representation $(\theta,N)$. 
\end{remark}
\begin{remark}
By Theorem \ref{th:weylquant}  the Weyl operator quantizing $\alpha$ in the representation $(\theta,N)$ depends only on the values of $\alpha$ on $L(\theta,N)$, namely on
$$
\alpha_{\theta,N}(r,s):=\alpha \left(\frac{r}{2N}+\frac{\theta_1}{N}, \frac{s}{2N}+\frac{\theta_2}{N} \right), \quad r,s = 0, \dots, 2N-1.
$$
 This means that the actual object that is quantized is the sampling $\alpha_{\theta,N}$ of the function $\alpha$ on  $L(\theta,N)$ (i.e. a $2N \times 2N$ matrix).
\end{remark}
We can now determine the most general class of functions on $\T^2$ admitting  Weyl quantization and analyze the non-injectivity of this map.
\begin{definition}\label{samplingoperator}
Let $\mathcal{F}(\T^2)$  denote the space of functions from $\T^2$ to $\C$.  The $(\theta,N)$-sampling operator $\mu_{\theta,N}: \mathcal{F}(\T^2) \to M_{2N}(\C)$ is defined as follows:
\begin{equation}
\mu_{\theta,N}(\alpha)=\alpha_{\theta,N}, \nonumber
\end{equation} 
where $\alpha_{\theta,N}(r,s)=\alpha \left(\frac{r}{2N}+\frac{\theta_1}{N}, \frac{s}{2N}+\frac{\theta_2}{N} \right)$, $r,s=0, \dots, 2N-1$.
\end{definition}
Then, clearly:
\begin{theorem}\label{thm:weylopegeneric}
Let  $\alpha\in\mathcal{F}(\T^2)$. Then  $\alpha$ admits Weyl quantization $\textrm{Op}^W_{\theta,N}(\alpha)$,  the $N \times N$ matrix with entries given by (\ref{matrixcoeffWeylquant}),  if and only if it admits a finite sampling  $\mu_{\theta,N}(\alpha)$ on $L(\theta,N)=\left\{\left(\frac{j}{2N}+\frac{\theta_1}{ N}, \frac{k}{2N}+\frac{\theta_2}{ N} \right): j,k \in \Z_{2N}\right\} \subset \T^2$.
\end{theorem}
Theorems \ref{th:weylquant}, \ref{thm:weylopegeneric} and Definition \ref{samplingoperator}  yield immediately the following Corollary  specifying  the contents of  Remark \ref{remarknotinjective}.
\begin{corollary}\label{corollarynotinjective}
Let $\alpha, \beta \in \mathcal{F}(\T^2)$ such that $\mu_{\theta,N}(\alpha)=\mu_{\theta,N}(\beta)$. Then $\textrm{Op}^W_{\theta,N}(\alpha)=\textrm{Op}^W_{\theta,N}(\beta)$.
\end{corollary}
\begin{remark}
By Corollary \ref{corollarynotinjective},  two functions taking the same values on  $L(\theta,N)$ generate the same Weyl operator in the representation $(\theta,N)$. The converse is not true, namely  functions assuming different values on $L(\theta,N)$ can generate the same Weyl operator in the representation $(\theta,N)$,  as shown in Theorem \ref{th:characteq}.
\end{remark}

To identify all  functions generating the same Weyl operator,  let us introduce an equivalence relation in the function space ${\mathcal F}(\T^2)$.
\begin{definition}
Let  $\alpha, \alpha' \in {\mathcal F}(\T^2)$. We say that $\alpha$ is equivalent to $\alpha^\prime$ in the representation $(\theta,N)$, and  write $\alpha \doteq_{\theta,N} \alpha'$,  if they generate the same Weyl operator, i.e. 
\begin{equation}\label{equivalencerel2}
\alpha \doteq_{\theta,N} \alpha' \iff  \textrm{Op}_{\theta,N}^W(\alpha)=\textrm{Op}_{\theta,N}^W(\alpha').
\end{equation}
\end{definition}

We  now  characterize the equivalence relation $\doteq_{\theta,N}$ by studying the kernel of the Weyl quantization procedure. To this end we introduce the operator 
\begin{equation}\label{def:delta}
\Delta: M_{2N}(\C) \to M_N(\C)
\end{equation}
defined as follows, for all $A=(A_{r,s})_{r,s=0,\dots,2N-1} \in M_{2N}(\C)$:
\begin{eqnarray}
&&
\Delta(A)=(\Delta(A)_{j,k})_{j,k=0, \dots, N-1} \nonumber
\\
&&
\Delta(A)_{j,k}:=A_{j,k}+(-1)^{k}A_{j+N,k}+(-1)^jA_{j,k+N}+(-1)^{j+k+N}A_{j+N,k+N}. \nonumber
\end{eqnarray}
By using the above definition one can prove the following theorem which is a central result of this paper.
 \begin{theorem}\label{th:characteq}
Let  $\alpha, \alpha' :\T^2 \to \C$ two functions on $\T^2$, then:
\begin{itemize}
\item [(i)] $\alpha \doteq_{\theta,N} \alpha'$ if and only if $\Delta(\mu_{\theta,N}(\alpha))=\Delta(\mu_{\theta,N}(\alpha'))$.
\item [(ii)] If $A=\textrm{Op}^W_{\theta,N}(\alpha)$ and $\mathcal{A}=\Delta(\mu_{\theta,N}(\alpha)) \in M_N(\C)$, then
\begin{equation}
A_{m,l}=\frac{1}{2N}\sum_{s=0}^{N-1}\mathcal{A}_{m+l,s} \;e^{\frac{\pi i s(m-l)}{N}}, \nonumber
\end{equation}
for all $m,l =0, \dots, N-1$, where the result of $m+l$ is modulo $N$.
\end{itemize}
\end{theorem}
\proof We start with the proof of (i) and denote by 
$$
x(r,s)=\alpha_{\theta,N}(r,s)-\alpha'_{\theta,N}(r,s)
$$
for all $r,s \in \Z_{2N}$. Since $\textrm{Op}_{\theta,N}^W(\alpha)=\textrm{Op}_{\theta,N}^W(\alpha')$, by (\ref{matrixcoeffWeylquant}) we have that for all $n,j = 0,\dots, N-1$
\begin{equation}
\sum_{m=0}^{2N-1}\left(x(j+n,m)+(-1)^mx(j+n+N,m)\right)e^{-\frac{\pi i im (j-n)}{N}}=0. \nonumber
\end{equation}
Introduce the  change of variables: 
$$
\left\{ \begin{array}{c} r=j+n \\
s= j-n
\end{array} \right.\qquad \textrm{(mod $2N$)}.
$$
It follows that
$$
\left\{ \begin{array}{c} 2j=r+s \\
2n= r-s
\end{array} \right.\qquad \textrm{(mod $2N$)},
$$
thus $r$ and $s$ must be both even or both odd.
We have only two cases: $(r,s)=(2r',2s')$ or $(r,s)=(2r'+1,2s'+1)$, for some $r',s' =0, \dots, N-1$. Thus in the first case we obtain: 
\begin{equation}\label{primocaso}
\sum_{m=0}^{2N-1}\left(x(2r',m)+(-1)^mx(2r'+N,m)\right)e^{-\frac{2\pi i im s'}{N}}=0,
\end{equation}
and in the second one:
\begin{equation}\label{secondocaso}
\sum_{m=0}^{2N-1}\left(x(2r'+1,m)+(-1)^mx(2r'+1+N,m)\right)e^{-\frac{2\pi i ims'}{N}}e^{-\frac{\pi i im}{N}}=0,
\end{equation}
for  $r',s'=0,\dots,N-1$. Now we multiply (\ref{primocaso}) and (\ref{secondocaso}) by $e^{\frac{2\pi i is'k}{N}}$,  with $k =0, \dots,N-1$, and sum with respect to $s'$ obtaining:
$$
\sum_{s'=0}^{N-1}e^{\frac{2\pi i is'k}{N}}\sum_{m=0}^{2N-1}\left(x(2r',m)+(-1)^mx(2r'+N,m)\right)e^{-\frac{2\pi i im s'}{N}}=0
$$
for the first case and 
$$
 \sum_{s'=0}^{N-1}e^{\frac{2\pi i is'k}{N}}\sum_{m=0}^{2N-1}\left(x(2r'+1,m)+(-1)^mx(2r'+1+N,m)\right)e^{-\frac{2\pi i ims'}{N}}e^{-\frac{\pi i im}{N}}=0
$$
for the second one.
Thus, since 
$$
 \sum_{s'=0}^{N-1}e^{\frac{2\pi i is'(k-m)}{N}}=N\delta^{(N)}_{m,k},
$$ 
we obtain that, for all $r',k =0, \dots,N-1$
$$
x(2r',k)+(-1)^kx(2r'+N,k)+x(2r',k)+(-1)^{k+N}x(2r'+N,k+N)=0
$$
and
$$
x(2r'+1,k)+(-1)^kx(2r'+1+N,k)-x(2r'+1,k)-(-1)^{k+N}x(2r'+1+N,k+N)=0.
$$
Therefore, for all $j,k =0, \dots,N-1$
$$
x(j,k)+(-1)^kx(j+N,k)+(-1)^jx(j,k+N)+(-1)^{j+k+N}x(j+N,k+N)=0
$$
and this concludes the proof of (i).

We now prove assertion (ii). We formally define for all $j,k=0,\dots,N-1$:
\begin{equation}
\mathcal{A}_{j+N,k}:=(-1)^k\mathcal{A}_{j,k}, \; \mathcal{A}_{j,k+N}:=(-1)^j\mathcal{A}_{j,k},\; \mathcal{A}_{j+N,k+N}:=(-1)^{j+k+N}\mathcal{A}_{j,k}. \nonumber
\end{equation}
By (\ref{matrixcoeffWeylquant}), a direct computation shows that for all $r,s=0,\dots,2N-1$:
\begin{equation}
\mathcal{A}_{r,s}=2 \sum_{j=0}^{N-1}A_{j,r-j}\;e^{-\frac{\pi i s(2j-r)}{N}}. \nonumber
\end{equation}
Therefore for all $k=0,\dots,2N-1$:
\begin{equation}\label{eqn:intermediateinversion}
\sum_{s=0}^{2N-1}\mathcal{A}_{r,s} e^{\frac{\pi i sk}{N}}=2 \sum_{j=0}^{N-1}A_{j,r-j}\sum_{s=0}^{2N-1}e^{-\frac{\pi i s(2j-r)}{N}}=4N\sum_{j=0}^{N-1}A_{j,r-j}\delta^{(2N)}_{2j,r+k}.
\end{equation}
The modular equation $2j=r+k$ (mod $2N$) has solution if and only if $r,k$ have the same parity, i.e. $(r,k)=(2r',2k')$ or $(r,k)=(2r'+1,2k'+1)$, for some $r',k'=0, \dots, N-1$. Therefore, by (\ref{eqn:intermediateinversion}) it results that
\begin{equation}
A_{r'+k',r'-k'}= \frac{1}{4N}\sum_{s=0}^{2N-1}\mathcal{A}_{2r',s} e^{\frac{\pi i s(2k')}{N}}, \quad \textrm{if  $(r,k)=(2r',2k')$} \nonumber
\end{equation} 
and
\begin{equation}
A_{r'+k'+1,r'-k'}= \frac{1}{4N}\sum_{s=0}^{2N-1}\mathcal{A}_{2r'+1,s} e^{\frac{\pi i s(2k'+1)}{N}}, \quad \textrm{if  $(r,k)=(2r'+1,2k'+1)$}  \nonumber
\end{equation} 
Now we introduce the  change of variables: 
$$
\left\{ \begin{array}{c} m=r'+k' \\
l= r'-k'
\end{array} \right.\qquad \textrm{(mod $N$)}
$$
for the first case and 
$$
\left\{ \begin{array}{c} m=r'+k'+1 \\
l= r'-k'
\end{array} \right.\qquad \textrm{(mod $N$)}
$$
for the second one. In both cases it results that
\begin{eqnarray}
A_{m,l}&=& \frac{1}{4N}\sum_{s=0}^{2N-1}\mathcal{A}_{m+l,s} \;e^{\frac{\pi i s(m-l)}{N}} \nonumber \\
&=& \frac{1}{4N}\sum_{s=0}^{N-1}\left( \mathcal{A}_{m+l,s} \;e^{\frac{\pi i s(m-l)}{N}} +\mathcal{A}_{m+l,s+N} \;e^{\frac{\pi i (s+N)(m-l)}{N}} \right) \nonumber \\
&=& \frac{1}{4N}\sum_{s=0}^{N-1}\left( \mathcal{A}_{m+l,s} \;e^{\frac{\pi i s(m-l)}{N}} +(-1)^{m+l}(-1)^{m-l}\mathcal{A}_{m+l,s} \;e^{\frac{\pi i s(m-l)}{N}} \right) \nonumber \\
&=& \frac{1}{2N}\sum_{s=0}^{N-1} \mathcal{A}_{m+l,s} \;e^{\frac{\pi i s(m-l)}{N}} \nonumber
\end{eqnarray}
and this concludes the proof. \qed
\begin{remark}
The Weyl quantization in the representation $(\theta,N)$ is defined for any function $\alpha\in {\mathcal F}(\T^2)$  given its sampling on  $L(\theta,N)$, namely the matrix 
$$
\mu_{\theta,N}(\alpha)=\{\alpha_{\theta,N}(r,s)\}_{r,s=0,\dots, 2N-1}\in M_{2N}(\C).
$$ 
Actually Theorem \ref{th:characteq} shows that the quantization is a linear correspondence between $M_{2N}(\C)$ (the space of all possible values $\{\alpha_{\theta,N}(r,s)\}_{r,s=0,\dots, 2N-1}$  and $M_N(\C)$ (the space of the quantum operators), namely
$$
M_{2N}(\C) \stackrel{\textrm{Op}^W_{\theta,N}}{\longrightarrow} M_N(\C).
$$
As  explained this correspondence is not injective and in order to obtain a bijection, we have to consider the appropriate quotient space. More precisely:
\end{remark}
\begin{corollary}
The Weyl quantization is a linear bijection between the quotient space  $M_{2N}(\C)/\textrm{Ker} \Delta$ and  $M_{N}(\C)$, where $\textrm{Ker} \Delta$ denotes the kernel of the operator $\Delta$ defined in (\ref{def:delta}).
\end{corollary}
\proof The proof of injectivity follows by Theorem \ref{th:characteq}. Moreover, since the dimension of $\textrm{Ker} \Delta$ is $3N^2$, it results that  $M_{2N}(\C)/\textrm{Ker} \Delta$ and $M_{N}(\C)$ have the same dimension $N^2$ so the the linear map is a bijection between the two spaces.   \qed
\begin{remark}
Assertion (i) of Theorem \ref{th:characteq} immediately entails  that functions assuming different values on  $L(\theta,N)$ may admit the same Weyl quantization. Moreover assertion (ii) shows how the Weyl operator explicitly depends on the equivalence class of the relation $\doteq_{\theta,N}$.
\end{remark}

\subsection{Moyal product and Moyal brackets}
The Moyal product represents the symbol of  the product of non-commuting operators as a deformed product of ordinary functions.  
\begin{definition} \label{defn: Moyal pr}
Let $\alpha, \beta\in \mathcal{F}(\T^2)$. The Moyal product $\alpha \; \sharp \; \beta$ is  (up to equivalence) the function on $\T^2$ generating the operator product  $\textrm{Op}^W_{\theta,N}(\alpha)\textrm{Op}^W_{\theta,N}(\beta)$ through Weyl quantization in the representation $(\theta,N)$. 
\end{definition}
\begin{theorem}
\label{th:moyal product}
Let $\alpha, \beta\in C^1(\T^2;\C)$. Then $\alpha \; \sharp \; \beta : \T^2 \to \C$ is given by
\begin{eqnarray}
\label{alphasharpbeta}
&&
(\alpha \; \sharp \; \beta)(x,p)=
\\
\nonumber
&&
=
\tfrac{1}{(2N)^2}\sum_{r_1,r_2,s_1,s_2=0}^{2N-1} \alpha \left( x+\tfrac{r_1}{2N},p+\tfrac{s_1}{2N} \right) \beta \left(x+\tfrac{r_2}{2N},p+\tfrac{s_2}{2N} \right)e^{\frac{\pi i}{N}(r_1s_2-r_2s_1)}.
\end{eqnarray}
\end{theorem}
\proof
By the Fourier inversion formula $\alpha$ and $\beta$: 
\begin{equation}\label{eq:fourierinversionalpha}
\alpha(x,p)=\sum_{n_1, n_2\in \Z}\hat{\alpha}(n_1, n_2)e^{2\pi i (n_1 x+n_2p)}  \nonumber
\end{equation}
and
\begin{equation}\label{eq:fourierinversionbeta}
\beta(x,p)=\sum_{m_1, m_2\in \Z}\hat{\beta}(m_1, m_2)e^{2\pi i (m_1 x+m_2p)}, \nonumber
\end{equation}
where the series converge uniformly.
From (\ref{def:Weyl op}) we have that
\begin{eqnarray}\label{moyal1}
&& \textrm{Op}^W_{\theta,N}(\alpha)\textrm{Op}^W_{\theta,N}(\beta) \nonumber \\
&=&  \mathop{\sum_{n_1, n_2\in \Z}}_{m_1, m_2\in \Z}   \hat{\alpha}(n_1, n_2)\hat{\beta}(m_1, m_2) T_{\theta,N}(n_1,n_2)T_{\theta,N}(m_1,m_2) \nonumber \\
&=&  \mathop{\sum_{n_1, n_2\in \Z}}_{m_1, m_2\in \Z} \hat{\alpha}(n_1, n_2)\hat{\beta}(m_1, m_2)e^{-\frac{i\pi}{N}(n_1m_2-n_2m_1)} T_{\theta,N}(n_1+m_1,n_2+m_2) \nonumber  \\
&=& \mathop{\sum_{n_1, n_2\in \Z}}_{j_1, j_2\in \Z}\hat{\alpha}(n_1, n_2)\hat{\beta}(j_1-n_1, j_2-n_2)e^{-\frac{i\pi}{N}(n_1j_2-n_2j_1)} T_{\theta,N}(j_1,j_2),\nonumber \\
&=& \sum_{j_1, j_2\in \Z}\hat{\gamma}(j_1, j_2)T_{\theta,N}(j_1,j_2)\nonumber 
\end{eqnarray}
where we used property 2 of Proposition~\ref{prop repr T} in the Appendix, and 
\begin{equation}
\hat{\gamma}(j_1, j_2):=\sum_{n_1, n_2\in \Z}\hat{\alpha}(n_1, n_2)\hat{\beta}(j_1-n_1, j_2-n_2)e^{-\frac{\pi i}{N}(n_1j_2-n_2j_1)} .\nonumber
\end{equation}
It is easy to prove that
\begin{equation}\label{hatgammaj1j2}
\hat{\gamma}(j_1, j_2)=\int_{\T^2} \alpha\left( x'-\frac{j_2}{2N}, p'-\frac{j_1}{2N}\right) \beta\left(x',p' \right) e^{2\pi i(x'j_1-p'j_2)}\; \mathrm{d}x' \; \mathrm{d}p' . 
\end{equation}
By definition:
\begin{equation}
\gamma(x,p)=\sum_{j_1, j_2\in \Z}\hat{\gamma}(j_1, j_2)e^{2\pi i (j_1 x+j_2p)} =\alpha \; \sharp \; \beta(x,p). \nonumber
\end{equation}
From (\ref{hatgammaj1j2}) it follows that  
\begin{eqnarray}
\label{moyprod}
&&
\alpha \; \sharp \; \beta(x,p)= \nonumber
\\
&&
\nonumber
\frac{1}{(2N)^2} \sum_{r_1,r_2,s_1,s_2=0}^{2N-1} \alpha \left( x+\tfrac{r_1}{2N},p+\tfrac{s_1}{2N} \right) \beta \left(x+\tfrac{r_2}{2N},p+\tfrac{s_2}{2N} \right)e^{\frac{\pi i}{N}(r_1s_2-r_2s_1)}. \nonumber
\end{eqnarray}
This concludes the proof. \qed

Out of the Moyal product of two functions we can  define their Moyal brackets.
\begin{definition} \label{defn: Moyal br}
Let $\alpha, \beta \in \mathcal{F}(\T^2)$. The Moyal brackets $\{\alpha , \beta\}_{\sharp}$ is the function (up to equivalence)  on $\T^2$ having the commutator  $[\textrm{Op}^W_{\theta,N}(\alpha),\textrm{Op}^W_{\theta,N}(\beta)]$ as Weyl operator in the representation $(\theta,N)$. 
\end{definition}
\begin{theorem}\label{Th:Moyal br}
Let $\alpha, \beta\in C^1( \T^2; \C)$. Then $\{\alpha , \beta\}_{\sharp} : \T^2 \to \C$ is given by
\begin{eqnarray}
\label{MoyalBracket}
&&
\{\alpha, \beta \}_{\sharp}(x,p) 
\\
&&
\nonumber
= \tfrac{2i}{(2N)^2}  \mathop{\sum_{r_1,r_2,}^{2N-1}}_{s_1,s_2=0} \alpha\left( x+\tfrac{r_1}{2N},p+\tfrac{s_1}{2N} \right) \beta\left( x+\tfrac{r_2}{2N},p+\tfrac{s_2}{2N} \right)\sin\left(\tfrac{\pi}{N} (r_1s_s-r_2s_1)\right). 
\end{eqnarray}
\end{theorem}
\proof
The proof is an elementary consequence of Theorem \ref{th:moyal product}. \qed

\begin{remark}
Theorems \ref{th:moyal product} - \ref{Th:Moyal br} show that the Moyal product and the Moyal brackets of two symbols are independent of $\theta$, they only depend on $N$.
\end{remark}
\subsection{Quantum dynamics in phase space: evolution of symbols on the torus}  \label{Quantum dynamics in phase space: evolution of symbols on the torus}

Let $\mathscr{H}: \T^2 \to \R$ be a function representing the \emph{classical Hamiltonian} and let 
$$
H_{\theta,N}=\textrm{Op}_{\theta,N}^W(\mathscr{H})
$$
be its Weyl quantization in the representation $(\theta,N)$. We consider a  fun\-ction $\alpha: \T^2 \to \C$ admitting finite sampling on $L(\theta,N)$  and its Weyl quantization $A_{\theta,N}=\textrm{Op}_{\theta,N}^W(\alpha)$ in the representation $(\theta,N)$. We can define the one-parameter group 
\begin{equation}
t \in \R \mapsto U(t)=e^{-2\pi i N t H_{\theta,N}} ,\nonumber
\end{equation}
and the operator $A_{\theta,N}(t)=U(-t)A_{\theta,N}U(t)$, $t \in \R$. An elementary computation shows that $t \in \R \mapsto A_{\theta,N}(t)$ solves the Heisenberg equation
\begin{equation} \nonumber
\left\{ \begin{array}{c}\displaystyle
\frac{i}{2\pi N} \frac{\mathrm{d}A(t)}{\mathrm{d}t}=\displaystyle[H_{\theta,N},A(t)], \\
\\
\displaystyle A(0)= \displaystyle A_{\theta,N}.
\end{array} \right.
\end{equation}
The  evolution equation for the symbol of the operator $A_{\theta,N}(t)$ is obtained via the  Moyal brackets in Definition \ref{defn: Moyal br}. Let $\alpha_{\theta,N}(t)$ be the symbol of $A_{\theta,N}(t)$.  Then $\alpha_{\theta,N}(t)$ is the solution of the following equation on the torus
\begin{equation}\label{MoyalequationN}
\left\{ \begin{array}{c} \displaystyle
\frac{i}{2\pi N} \frac{\mathrm{d}\alpha(t)}{\mathrm{d}t}=\{\mathscr{H},\alpha(t)\}_{\sharp}, \\
\\
\alpha(0)=\alpha.
\end{array} \right.
\end{equation}
\begin{remark}
From (\ref{MoyalBracket}), proceeding as  is  the case of the standard semiclassical calculus in $\R^2$ (see e.g \cite{folland,Martinez,Robert}) it is not difficult to prove that
\begin{equation}\label{asymptoticsmoyalb1}
\frac{2\pi N}{i} \{\alpha,\beta\}_\sharp(x,p) = \{\alpha,\beta\}(x,p)+O\left( \left(\frac{1}{2\pi N}\right)^2 \right),
\end{equation}
for all $\alpha, \beta \in C^\infty(\T^2; \C)$, where $ \{\alpha,\beta\}$ are the Poisson brackets of $\alpha$ and $\beta$, namely
\begin{equation} \nonumber
 \{\alpha,\beta\}=\frac{\partial\alpha}{\partial x}\frac{\partial\beta}{\partial p}-\frac{\partial\alpha}{\partial p}\frac{\partial\beta}{\partial x}.
\end{equation}
This implies that in the $N \to \infty$ limit of (\ref{MoyalequationN}) one recovers the classical equation of motion of the Hamiltonian dynamics 
\begin{equation}\label{poissonequation}
\left\{ \begin{array}{c}\displaystyle
\frac{\mathrm{d}\alpha(t)}{\mathrm{d}t}=\{\mathscr{H},\alpha(t)\} \\
\alpha(0)=\alpha.
\end{array} \right.
\end{equation}
In this sense (\ref{MoyalequationN}) represents the quantization of the generic classical dynamical system on the torus given by (\ref{poissonequation}).
\end{remark}

\section{Wigner transform on the torus}
In this section we construct the Wigner transform on $\T^2$ starting from the representation of the discrete Heisenberg group. We follow the approach in \cite{DEGI}, with a suitable modification of the definition, that takes into account the fact that the representation $T_{\theta,N}$ is not $N$-periodic. Then we compute explicitly the Wigner transform and determine its support. 
\begin{definition}\label{defin:Wigner1}
Let $\psi, \varphi \in \C^N$, we
define the Fourier-Wigner transform of $\psi$ and $\varphi$ in the representation $(\theta,N)$ as $V_{\theta,N}(\psi,\varphi):\Z^2 \to \C$ such that 
\begin{equation}\label{def:fourierW}
V_{\theta,N}(\psi,\varphi)(n_1,n_2)=\langle \psi, T_{\theta,N}(n_1, n_2) \varphi \rangle, \qquad \textrm{for all $(n_1,n_2) \in \Z^2$},
\end{equation}
where $\langle \cdot , \cdot \rangle$ denotes the scalar product on  $\C^N$.
We define the Wigner transform of $\psi$ and $\varphi$ in the representation $(\theta,N)$ as the distribution on $\T^2$ defined by
\begin{equation}\label{defin:Wigner}
W_{\theta,N}(\psi,\varphi)(x,p)=\sum_{n_1, n_2, \in \Z}V_{\theta,N}(\psi,\varphi)(n_1,n_2)e^{-2\pi i (n_1x+ n_2p)},
\end{equation}
where the Fourier series converges in the sense of distribution (since $V_{\theta,N}(\psi,\varphi)$ is uniformly bounded).
\end{definition}

Before computing the explicit formula for the Wigner transform $W_{\theta,N}(\psi,\varphi)$ we present some of its basic properties.  The following Proposition is the analog of Proposition 3.2 of \cite{DEGI} for the modified definition of the Wigner transform on $\T^2$ in Eq. (\ref{defin:Wigner}). We introduce some notations: if $D$ is a distribution on $\T^2$ ($D \in \mathscr{D}'(\T^2)$) and $\phi $ is a test function on $\T^2$ ($\phi \in \mathscr{D}(\T^2)$), we denote with $\int_{\T^2}D(x,p)\phi(x,p)\; \mathrm{d}x \; \mathrm{d}p $ the action of $D$ on $\phi$ and with $\delta_{\Z}$ the Dirac delta on $\T^2$, namely 
\begin{equation}
 \delta_{\Z}(x)=\sum_{m \in \Z}\delta(x-m).
\end{equation}
\begin{proposition}\label{propT}
Let $\psi, \varphi  \in \C^N$,  then
\begin{enumerate}
\item
\begin{equation}
\int_{\T^2}W_{\theta,N}(\psi,\varphi)(x,p)\; \mathrm{d}x \; \mathrm{d}p = \langle \psi, \varphi \rangle;
\end{equation}
\item 
\begin{equation}\label{supp1marg}
\int_{\T^1 }W_{\theta,N}(\psi,\varphi)(x,p)\; \mathrm{d}p = \sum_{j=0}^{N-1}\overline{\psi_j}\varphi_j \delta_{\Z}\left( x- \frac{j+\theta_1}{N}\right);
\end{equation}
\item 
\begin{equation}\label{supp2marg}
\int_{\T^1 }W_{\theta,N}(\psi,\varphi)(x,p)\; \mathrm{d}x = \frac{1}{N}\sum_{j=0}^{N-1}\overline{\hat{\psi}_j}\hat{\varphi}_j \delta_{\Z}\left(p-\frac{j+\theta_2}{N} \right),
\end{equation}
where $\hat{\psi}_j$, $\hat{\varphi}_j $, $j=0, \dots, N-1$ are the discrete Fourier coefficients of $\psi$ and $\varphi$: 
\begin{equation}
\hat{\psi}_j=\sum_{m=0}^{N-1}\psi_m e^{-\frac{2\pi i mj}{N}} \qquad \textrm{and} \qquad \hat{\varphi}_j=\sum_{m=0}^{N-1}\varphi_m e^{-\frac{2\pi i mj}{N}}. \nonumber
\end{equation}
\item
\begin{equation}
\overline{W_{\theta,N}(\varphi,\psi)}=W_{\theta,N}(\psi,\varphi).
\end{equation}
In particular, if  for $\varphi=\psi$ we set $W_{\theta,N}\psi:=W_{\theta,N}(\psi,\psi)$,  then $W_{\theta,N}\psi$ is real.
\item
For any $\alpha\in C^1(\T^2;\C)$ we have:
\begin{equation}
\langle \psi, \textrm{Op}_{\theta,N}^W(\alpha) \varphi \rangle =\int_{\T^2}\alpha(x,p) W_{\theta,N}(\psi,\varphi)(x,p)\; \mathrm{d}x \; \mathrm{d}p . 
\end{equation}
\end{enumerate}
\end{proposition}
\proof 
First, notice that a simple calculation yields:
\begin{eqnarray} 
V_{\theta,N}(\psi,\varphi)(n_1,n_2) &=& \sum_{j,l =0}^{N-1}\overline{\psi}_j \varphi_l e^{-\frac{\pi in_1 n_2}{N}}e^{\frac{2\pi in_1}{N}\left(l+\theta_1 \right)} e^{\frac{2 \pi i\theta_2n_2}{N}}\delta^{(N)}_{j,l-n_2} \nonumber \\
&=&e^{\frac{2 \pi i(n_1\theta_1+n_2\theta_2)}{N}} e^{-\frac{\pi in_1 n_2}{N}} \sum_{l =0}^{N-1}\overline{\psi}_{l-n_2} \varphi_l e^{\frac{2\pi i  ln_1}{N}}, \nonumber
\end{eqnarray}
where $\delta^{(N)}_{j,m}$ is the Kronecker delta on $\Z_N$.

Let us begin with the proof of assertion 1:
\begin{eqnarray}\label{proofi}
&& \int_{\T^2}W_{\theta,N}(\psi,\varphi)(x,p)\; \mathrm{d}x \; \mathrm{d}p \nonumber \\
&=&\int_{\T^2} \sum_{n_1, n_2, \in \Z}V_{\theta,N}(\psi,\varphi)(n_1,n_2)e^{-2\pi i (n_1x+ n_2p)}\; \mathrm{d}x \; \mathrm{d}p  \nonumber \\
&=& \sum_{n_1, n_2, \in \Z}\left(\int_{\T^1}e^{-2\pi in_1x}\; \mathrm{d}x \right)\left(\int_{\T^1}e^{-2\pi in_2p}\;\mathrm{d}p \right)V_{\theta,N}(\psi,\varphi)(n_1,n_2)  \nonumber  \\
&=& V_{\theta,N}(\psi,\varphi)(0,0)=\langle \psi, \varphi \rangle. \nonumber
\end{eqnarray}

We now prove  assertion 2. First notice that
\begin{eqnarray}\label{primo marg}
&& \int_{\T^1}W_{\theta,N}(\psi,\varphi)(x,p)\; \mathrm{d}p \nonumber \\
&=& \sum_{n_1, n_2 \in \Z}\left(\int_{\T^1}e^{-2\pi in_2p}\;\mathrm{d}p\right)V_{\theta,N}(\psi,\varphi)(n_1,n_2)e^{-2\pi i n_1x}   \nonumber \\
&=&  \sum_{n_1 \in \Z}V_{\theta,N}(\psi,\varphi)(n_1,0)e^{-2\pi in_1x} \nonumber \\ 
&=& \sum_{n_1\in \Z} e^{\frac{ 2\pi i n_1 \theta_1}{N}}\sum_{l=0}^{N-1}\overline{\psi}_{l} \varphi_l e^{\frac{2\pi in_1 l}{N}}e^{-2\pi in_1 x}. \nonumber
\end{eqnarray}
Then, applying the Poisson summation formula (\ref{Poisson Formula}) in the summation with respect to $n_1 \in \Z$, we obtain
\begin{eqnarray}
\int_{\T^1}W_{\theta,N}(\psi,\varphi)(x,p)\; \mathrm{d}p
&=&\sum_{l=0}^{N-1}\overline{\psi}_{l} \varphi_l \sum_{n_1\in \Z} \delta \left(x-\frac{l+ \theta_1}{N}-n_1 \right) \nonumber  \\
&=&\sum_{l=0}^{N-1}\overline{\psi}_{l} \varphi_l \sum_{n_1\in \Z} \delta \left(x-\frac{l +n_1N +\theta_1}{N}\right), \nonumber
\end{eqnarray}
as desired.

The proof of assertion  3 is quite similar: we start with the Fourier inversion formula that gives the following identities
\begin{equation}\label{discrete fourier inversion}
\psi_j=\frac{1}{N} \sum_{l=0}^{N-1}\hat{\psi}_l e^{\frac{2\pi  l j}{N}}, \qquad \varphi_j=\frac{1}{N} \sum_{l=0}^{N-1}\hat{\varphi}_l e^{\frac{2\pi  l j}{N}}, \; \textrm{for all $j =0, \dots, N-1$}.\nonumber
\end{equation}
We have that
\begin{equation}\label{secondo marg}
 \int_{\T^1}W_{\theta,N}(\psi,\varphi)(x,p)\; \mathrm{d}p 
= \sum_{n_2 \in \Z} V_{\theta,N}(\psi,\varphi)(0,n_2)e^{-2\pi in_2p},  \nonumber
\end{equation}
then using (\ref{discrete fourier inversion}) we get:
\begin{eqnarray}
&& \sum_{n_2 \in \Z} V_{\theta,N}(\psi,\varphi)(0,n_2)e^{-2\pi in_2p} \nonumber \\
 &=&  \frac{1}{N^2} \sum_{n_2\in \Z} e^{\frac{2 \pi i  n_2 \theta_2}{N}}\sum_{ j,l,m =0}^{N-1} \overline{\hat{\psi}_j}e^{-\frac{2\pi ij(l-n_2) }{N}} \hat{\varphi}_m e^{\frac{2\pi i ml}{N}}e^{-2\pi in_2 p}\nonumber \\
&=&\frac{1}{N^2} \sum_{ j,m =0}^{N-1} \sum_{n_2\in \Z} e^{-2\pi i n_2 \left( p-\frac{j +\theta_2 }{N}\right) } \overline{\hat{\psi}_j}\hat{\varphi}_m \sum_{l =0}^{N-1}e^{-\frac{2\pi il(j-m) }{N}}.  \nonumber
\end{eqnarray}
Using the Poisson formula (\ref{Poisson Formula}) in the summation with respect to $n_2 \in \Z$ and the  identity
\begin{equation}\label{discreteKdelta}
\sum_{k =0}^{N-1}e^{\frac{2 \pi i mk}{N}}=N \delta^{(N)}_{m,0}\nonumber
\end{equation}
in the summation with respect to  $l=0, \dots, N-1$ we finally  obtain:
\begin{eqnarray}
\int_{\T^1}W_{\theta,N}(\psi,\varphi)(x,p)\; \mathrm{d}p 
&=&\frac{1}{N} \sum_{j  =0}^{N-1}  \overline{\hat{\psi}_j} \hat{\varphi}_j \sum_{n_2\in \Z} \delta \left( p-\frac{j +n_2N+ \theta_2}{N} \right), \nonumber
\end{eqnarray}
as desired.

The proof of assertion 4 is a direct consequence of (\ref{defin:Wigner}).
Finally, the proof of assertion 5 is a direct consequence of (\ref{def:fourierW}) and (\ref{defin:Wigner}):
\begin{eqnarray}
\langle \psi, \textrm{Op}_{\theta,N}^W(\alpha) \varphi \rangle &=& \sum_{n_1, n_2 \in \Z} \hat{\alpha}(n_1 n_2)  \langle \psi, T_{\theta,N}(n_1,n_2) \varphi \rangle \nonumber \\
&=& \sum_{n_1, n_2 \in \Z}   \hat{\alpha}(n_1 n_2) V_{\theta,N}(\psi, \varphi)(n_1,n_2)   \nonumber \\
&=& \sum_{n_1, n_2 \in \Z}  V_{\theta,N}(\psi, \varphi)(n_1,n_2) \int_{\T^2} \alpha(x,p)e^{-2\pi i (x n_1+ p n_2)}\; \mathrm{d}x \; \mathrm{d}p  \nonumber \\
&=& \int_{\T^2} \alpha(x,p)W_{\theta,N}(\psi, \varphi)(x,p)\; \mathrm{d}x \; \mathrm{d}p  . \nonumber
\end{eqnarray}
\qed
\begin{remark}
Assertion 5 of Proposition \ref{propT} can be extended, in an appropriate sense, to any function  $\alpha:\T^2\to\C$ admitting finite sampling, as  shown in Proposition \ref{propTdisc} and in Remark \ref{extensionexpval} below.
\end{remark}

The next theorem yields an explicit formula for the Wigner transform, and shows that it is actually a signed measure on $\T^2$ with discrete support.
\begin{theorem}\label{th:explicitWigner}
Let $\psi, \varphi \in \C^N$,  then 
\begin{equation}\label{explicitWigner}
W_{\theta,N}(\psi,\varphi)= \sum_{r,s =0}^{2N-1}  \tilde{W}_N(\psi,\varphi)(r,s) \; \;\delta_{\Z} \left( x-\tfrac{r}{2N}-\tfrac{\theta_1}{N}\right)  \delta_{\Z} \left( p-\tfrac{s}{2N}-\tfrac{\theta_2}{N}\right),
\end{equation}
where 
\begin{equation}\label{def:tildeWigner}
\tilde{W}_N(\psi,\varphi)(r,s)=\frac{1}{2N}  \sum_{l \in \Z_N}\overline{\psi_{r-l}}\varphi_l e^{-\frac{\pi i}{N}(2l-r) s}, \qquad r,s =0, \dots, 2N-1.
\end{equation}
\end{theorem}
\proof
We start with the definition of $ W_{\theta,N}(\psi,\varphi)$ given in Eq. (\ref{defin:Wigner}):
\begin{eqnarray} \label{calc: Wignerstep1}
&& W_{\theta,N}(\psi,\varphi)(x,p)\nonumber \\
&=& \sum_{n_1,n_2 \in \Z} V_{\theta,N}(\psi,\varphi)(n_1,n_2)e^{-2\pi i (n_1 x+n_2 p)} \nonumber \\
                                                      &=&  \sum_{j_1,j_2 =0}^{2N-1}  \sum_{n_1,n_2  \in \Z} V_{\theta,N}(\psi,\varphi)(j_1+2Nn_1,j_2+2Nn_2)e^{-2\pi i (j_1 x+j_2 p)}e^{-4\pi iN (n_1 x+n_2 p)}  \nonumber \\
                                                      &=&    \sum_{j_1,j_2 =0}^{2N-1}  \sum_{n_1,n_2  \in \Z} e^{2 \pi i(2n_1 \theta_1+2n_2 \theta_2)}V_{\theta,N}(\psi,\varphi)(j_1,j_2)  e^{-2\pi i (j_1 x+j_2 p) }e^{-4\pi i N(n_1x+n_2p)}, \nonumber
\end{eqnarray}
where we used property 4 of Proposition~\ref{prop repr T} in the Appendix. Now, again by the Poisson formula  (\ref{Poisson Formula}), we have that
\begin{eqnarray}
&& \sum_{n_1,n_2  \in \Z} e^{-2\pi i \left[ n_1\left( 2Nx-2\theta_1\right)+n_2\left( 2Np-2\theta_2\right)  \right]} \nonumber \\
&&=\tfrac{1}{(2N)^2}\sum_{n_1,n_2 \in \Z}  \delta \left( x-\tfrac{\left( n_1+2\theta_1\right)}{2N}\right)  \delta \left( p-\tfrac{\left( n_2+2\theta_2\right)}{2N}\right), \nonumber
\end{eqnarray}
therefore,
\begin{eqnarray}
&& W_{\theta,N}(\psi,\varphi)(x,p) \nonumber \\
&& = \sum_{r,s =0}^{2N-1}  \tilde{W}_N(\psi,\varphi)(r,s) \sum_{m_1,m_2 \in \Z} \delta \left( x-\left(\tfrac{r+2Nm_1}{2N}+ \tfrac{\theta_1}{N}\right)\right)  \delta \left( p-\left(\tfrac{s+2Nm_2}{2N}+\tfrac{\theta_2}{ N} \right)\right) ,                            \nonumber
\end{eqnarray}
where
\begin{equation}
\tilde{W}_N(\psi,\varphi)(r,s)=\frac{1}{(2N)^2} \sum_{j_1,j_2 =0}^{2N-1} V_{\theta,N}(\psi,\varphi)(j_1,j_2)  e^{-\frac{2\pi i}{2N} \left[j_1 \left( 2\theta_1+r\right)+j_2 \left( 2\theta_2+s \right)\right]} . \nonumber
\end{equation}
Now we compute explicitly $\tilde{W}_N(\psi,\varphi)$: for $r,s =0, \dots , 2N-1$
\begin{eqnarray}\label{calc:Wignerstep2}
\tilde{W}_N(\psi,\varphi)(r,s) &=& \frac{1}{(2N)^2}  \mathop{\sum_{j_1,j_2 =0, \dots , 2N-1}}_{l =0,\dots, N-1}   \overline{\psi_{l-j_2}}\varphi_l e^{-\frac{\pi i j_1j_2}{N}}e^{\frac{\pi i j_1l}{N}} e^{-\frac{2\pi i}{2N} (j_1 r+j_2 s)} \nonumber\\
&=& \frac{1}{(2N)^2} \sum_{j_2=0}^{2N-1} \sum_{l =0}^{N-1}\overline{\psi_{l-j_2}}\varphi_l e^{-\frac{\pi i j_2s}{N}} \sum_{j_1=0}^{2N-1}e^{-\frac{2\pi i j_1(j_2+r-2l)}{2N}}\nonumber \\
&=& \frac{1}{2N} \sum_{l =0}^{N-1}\overline{\psi_{l-j_2}}\varphi_l e^{-\frac{\pi i j_2 s}{N}} \delta^{(2N)}_{j_2, 2l-r}\nonumber \\
&=&\frac{1}{2N}  \sum_{l =0}^{N-1}\overline{\psi_{r-l}}\varphi_l e^{-\frac{\pi i}{N}(2l-r) s}, \nonumber
\end{eqnarray}
and this concludes the proof.
\qed
Several remarks are in order. 
\begin{remark}
In classical mechanics the state of a system is characterized by a probability measure on the phase space, while the observables are described by real functions. The value of any observable in the state of the system is obtained by the pairing between the two objects. 
In the standard phase space formulation of quantum mechanics  (where the Hilbert space is $L^2(\R^n)$ and the phase space  $T^\ast \R^n \simeq  \R^n \times  \R^n$), the quantum state $\psi \in L^2(\R^n)$ is described by its Wigner transform, which is a function on the phase space that can be associated to an absolutely continuous measure.  The observables are functions on phase space as well.
If the phase space is the torus $\T^2$,  Definitions \ref{defin:Wigner} and \ref{def:Weyl op}   immediately yield  that the state is described by the Wigner transform, which is a distribution on $\T^2$  (signed measure), while the observables are represented by continuous function on $\T^2$, as  in the classical formalism.  Moreover the value of any observable in the given state of the system is obtained by the pairing described in 5 of Proposition \ref{propT}.
\end{remark}
\begin{remark}
The Wigner transform $W_{\theta,N}(\psi,\varphi)$ in (\ref{explicitWigner}) depends on the representation of the Heisenberg group, i.e. on $\theta$, only in its support, while $\tilde{W}_N(\psi, \varphi)$ is   independent of $\theta$. Moreover,  for all $r,s=0,\dots,2N-1$  
\begin{equation}
\tilde{W}_N(\psi,\varphi)(r+2N,s+2N)=\tilde{W}_N(\psi,\varphi)(r,s), \nonumber
\end{equation}
thus  $\tilde{W}_N(\psi,\varphi)$ is  actually defined on $\Z_{2N}\times \Z_{2N}$.
\end{remark}

\begin{remark}
It can be easily proved that the values of the functions $\tilde{W}_N$ and $V_{\theta,N}$ are related, as in the standard quantization on $\R^2$, by the following formula:
\begin{equation}\label{fourier-Wigner}
V_{\theta,N}(\psi,\varphi)(k,m)= 2N\, \tilde{W}_N(\tilde{\psi},\varphi)(m,-k) \; e^{\frac{2\pi i (k\theta_1+m\theta_2)}{N}},
\end{equation}
for all $k,m =0, \dots, 2N-1$, $\theta=(\theta_1, \theta_2)\in \T^2$, $\psi=\sum_j \psi_j u_j, \varphi =\sum_j \varphi_j u_j\in \C^N$, where 
$$
\tilde{\psi}=\sum_{j=0}^{N-1} \psi_{N-j}u_j.
$$
\end{remark}

\begin{remark}\label{remarksuppW}
Let $M_x[W_{\theta,N}(\psi,\varphi)]$ and $M_p[W_{\theta,N}(\psi,\varphi)]$ denote the two marginals of $W_{\theta,N}(\psi,\varphi)$, namely
\begin{equation}
M_x[W_{\theta,N}(\psi,\varphi)](x)=\int_{\T^1}W_{\theta,N}(\psi,\varphi)(x,p)\;  \mathrm{d}p, 
\end{equation}
and
\begin{equation}
M_p[W_{\theta,N}(\psi,\varphi)](p)=\int_{\T^1}W_{\theta,N}(\psi,\varphi)(x,p)\; \mathrm{d}x, 
\end{equation}
We can now  analyze the support of the Wigner transform:  (\ref{supp1marg}) and (\ref{supp2marg}) entail that  
$$
{\rm supp}\; M_x[W_{\theta,N}(\psi,\varphi)] \subset \mathscr{L}(\theta_1,N)=\left\{ \frac{j}{N}+\frac{\theta_1}{N}: j \in \Z_N  \right\}
$$
and  
$$
{\rm supp}\; M_p[W_{\theta,N}(\psi,\varphi)] \subset \mathscr{L}(\theta_2,N)=\left\{ \frac{j}{N}+\frac{\theta_2}{ N}: j \in \Z_N \right\}.
$$ 
Then one can naively expect that  ${\rm supp}\; W_{\theta,N}(\psi,\varphi) \subset\mathscr{L}(\theta_1,N) \times \mathscr{L}(\theta_2,N)$, but (\ref{explicitWigner}) shows that this is not the case!  We have indeed:
$$
{\rm supp}\; W_{\theta,N}(\psi,\varphi) \subset L(\theta,N)=\left\{ \left(\frac{r}{2N}+ \frac{\theta_1}{N}, \frac{s}{2N}+\frac{\theta_2}{ N} \right):r,s \in \Z_{2N} \right\}.
$$
 Clearly  $\mathscr{L}(\theta_1,N) \times \mathscr{L}(\theta_2,N)$ is a proper subset of $L(\theta,N)$: it corresponds to the  ``even pairs'', i.e. $\left\{\left(\frac{2j}{2N}+\frac{\theta_1}{N},\frac{2m}{2N}+\frac{\theta_2}{ N}\right): j,m \in \Z_N \right\}$.  However there is another lattice, the \emph{ghost} lattice,  that corresponds to all  other cases and cannot be ignored as shown below. The double lattice is there because the map $T_{\theta,N}$ defined in (\ref{defrepresteta}) is  $2N$-periodic (up to a phase factor) and not just $N$-periodic.  This  ``double'' periodicity is inherited by the Wigner transform. 
\end{remark}
\begin{remark}
The support of the Wigner transform has been noticed  in the original paper by Berry and Hannay \cite{berry}, and then revisited in \cite{Zak}, together with the following symmetries: 
\begin{equation}\label{propWigner1}
\tilde{W}_N(\psi, \varphi)(m+N,l)=(-1)^l\tilde{W}_N(\psi, \varphi)(m,l),  
\end{equation}
\begin{equation}\label{propWigner2}
\tilde{W}_N(\psi, \varphi)(m,l+N)=(-1)^m\tilde{W}_N(\psi, \varphi)(m,l),
\end{equation}
\begin{equation}\label{propWigner3}
\tilde{W}_N(\psi, \varphi)(m+N,l+N)=(-1)^{m+l+N}\tilde{W}_N(\psi, \varphi)(m,l),
\end{equation}
for any $m,l =0, \dots, N-1$.  The proof of (\ref{propWigner1}), (\ref{propWigner2}), (\ref{propWigner3}) is an immediate  consequence of (\ref{def:tildeWigner}). These symmetries derives from the fact that $\varphi$ and $\psi$ are characterized by $N$ complex numbers, so that the Wigner transform can assume  at most $N^2$ independent values. Therefore  the Wigner transform can be restricted e. g. to the proper $N \times N$ sub-lattice
\begin{equation}
I(\theta,N)=\left\{ \left( \frac{m}{2N} + \frac{\theta_1}{N}, \frac{l}{2N} + \frac{\theta_2}{N} \right): m,l =0, \dots,N-1 \right\},\nonumber
\end{equation}
the  \emph{independent lattice},  where it assumes the $N^2$ independent values. Note, however, that 
$I(\theta,N)$ differs from the cartesian product of the support of the two marginals of the Wigner transform, i.e. $I(\theta,N) \neq \mathscr{L}(\theta_1,N) \times \mathscr{L}(\theta_2,N)$, and that the values assumed on $\mathscr{L}(\theta_1,N) \times \mathscr{L}(\theta_2,N)$ are not independent.
\end{remark}
 Theorem \ref{th:explicitWigner}  immediately yields the following properties for the values of the Wigner transform, namely the analog of Proposition \ref{propT} for the map $\tilde{W}_N$.
\begin{proposition}\label{propTdisc}
Let $\psi=(\psi_j)_{j=0, \dots, N-1}, \varphi=(\varphi_{k})_{k=0, \dots, N-1} \in \C^N$.
Then:
\begin{enumerate}
\item
\begin{equation}
\sum_{r,s=0}^{2N-1}\tilde{W}_N(\psi,\varphi)(r,s)= \langle \psi, \varphi \rangle;
\end{equation}
\item 
\begin{equation}\label{supp1margdisc}
\sum_{s=0}^{2N-1}\tilde{W}_N(\psi,\varphi)(r,s)= \sum_{j=0}^{N-1}\overline{\psi_j}\varphi_j \delta^{(2N)}_{r,2j};
\end{equation}
\item 
\begin{equation}\label{supp2margdisc}
\sum_{r=0}^{2N-1}\tilde{W}_N(\psi,\varphi)(r,s)=  \frac{1}{N}\sum_{j=0}^{N-1}\overline{\hat{\psi}_j}\hat{\varphi}_j \delta^{(2N)}_{r,2j},
\end{equation}
where $\hat{\psi},\,\hat{\varphi}_j $, $j=0, \dots, N-1$ are the Fourier coefficients of $\psi$ and $\varphi$: 
\begin{equation}
\hat{\psi}_j=\sum_{m=0}^{N-1}\psi_m e^{-\frac{2\pi i mj}{N}} \qquad \textrm{and} \qquad \hat{\varphi}_j=\sum_{m=0}^{N-1}\varphi_m e^{-\frac{2\pi i mj}{N}}.
\end{equation}
\item
\begin{equation}
\overline{\tilde{W}_{N}(\varphi,\psi)}=\tilde{W}_{N}(\psi,\varphi),
\end{equation}
In particular, if for $\varphi=\psi$ we define $\tilde{W}_{N}\psi:=\tilde{W}_{N}(\psi,\psi)$, then  $\tilde{W}_{N}\psi$ is  real.
\item
For any function $\alpha\in{ \mathcal F}(\T^2)$ we have:
\begin{equation}
\langle \psi, \textrm{Op}_{\theta,N}^W(\alpha) \varphi \rangle =\sum_{r,s=0}^{2N-1}\alpha_{\theta,N}(r,s)\tilde{W}_N(\psi,\varphi)(r,s),
\end{equation}
where for all $r,s=0, \dots, 2N-1$
$$
\alpha_{\theta,N}(r,s):=\alpha\left( \frac{r}{2N}+\frac{\theta_1}{N}, \frac{s}{2N}+\frac{\theta_2}{N}\right),
$$
namely $\alpha_{\theta,N}=\mu_{\theta,N}(\alpha)$.
\end{enumerate}
\end{proposition}

\begin{remark}\label{extensionexpval}
Assertion 5 in Proposition \ref{propTdisc} is the extension of  assertion 5 in Proposition  \ref{propT} to any  function $\alpha\in \mathcal{F}(\T^2)$ having finite sampling on $L(\theta,N)$. The proof of the formula follows from a direct computation of $\langle \psi, \textrm{Op}_{\theta,N}^W(\alpha) \varphi \rangle$ using the matrix elements of $ \textrm{Op}_{\theta,N}^W(\alpha) $ in (\ref{matrixcoeffWeylquant}).
\end{remark}

We conclude this section defining  the Wigner transform of a linear operator acting on $\C^N$ which will be useful in the next section.
\begin{definition}
Given a linear operator  $F$ acting on $\C^N$, we
%\begin{equation}\label{def:psiphi}
%\psi=\sum_{j =0}^{N-1}\psi_{j}u_j  \qquad \textrm{and} \qquad \varphi=\sum_{l=0}^{N-1} \varphi_l u_l
%\end{equation}
define the Fourier-Wigner transform of $F$  in the representation $(\theta,N)$, $V_{\theta,N}(F):\Z^2 \to \C$ as
\begin{equation}
V_{\theta,N}(F)(n_1,n_2)=\textrm{tr} \left(T_{\theta,N}(n_1, n_2) F \right), \qquad \textrm{for all $ (n_1,n_2) \in \Z^2$},
\end{equation}
where $\textrm{tr}$ denotes the trace.
\newline
We define the Wigner transform of $F$ in the representation $(\theta,N)$ as the distribution on $\T^2$ defined by
\begin{equation}\label{def:Wignerop}
W_{\theta,N}(F)(x,p)=\sum_{n_1, n_2, \in \Z}V_{\theta,N}(F)(n_1,n_2)e^{-2\pi i (n_1x+ n_2p)},
\end{equation}
where the Fourier series converges in the sense of distribution, (since $V_{\theta,N}(F)$ is uniformly bounded).
\end{definition}
An immediate consequence of Theorem \ref{explicitWigner} is the following.
\begin{corollary}\label{Wigmatrix}
Let $F$ be  a linear operator in $\C^N$ represented by the matrix $(F_{j,k})_{j,k=0, \dots,N-1}$. Then: 
\begin{equation}\label{Wignertens1}
W_{\theta,N}(F)(x,p)= \sum_{r,s=0}^{2N-1} \tilde{W}_N(F)(r,s)\; \delta_{\Z} \left( x-\tfrac{r}{2N}-\tfrac{\theta_1}{N}\right)  \delta_{\Z} \left( p-\tfrac{s}{2N}-\tfrac{\theta_2}{N}\right) 
\end{equation}
where
\begin{equation}\label{Wignertens2}
\tilde{W}_N(F)(r,s)=\frac{1}{2N}  \sum_{l =0}^{N-1}F_{l, r-l}e^{-\frac{\pi i}{N}(2l-r) s}.
\end{equation}
Moreover for all $r,s=0,2N-1$
\begin{equation}\label{discWigF}
\tilde{W}_N(F)(r+2N,s+2N)=\tilde{W}_N(F)(r,s). 
\end{equation}
\end{corollary}
Therefore $\tilde{W}_N(F)$ is actually defined on $\Z_{2N} \times \Z_{2N}$. Finally:
\begin{corollary}\label{cor:symmetriesWF}
Let $F$ be  a linear operator in $\C^N$. Then: 
\begin{equation}\label{prop:Wignaone}
\tilde{W}_N(F)(m+N,l)=(-1)^l\tilde{W}_N(F)(m,l),
\end{equation}
\begin{equation}\label{prop:Wignatwo}
\tilde{W}_N(F)(m,l+N)=(-1)^m\tilde{W}_N(F)(m,l)  ,
\end{equation}
\begin{equation}\label{prop:Wignathree}
\tilde{W}_N(F)(m+N,l+N)=(-1)^{m+l+N}\tilde{W}_N(F)(m,l) , 
\end{equation}
for any $m,l =0 \dots, N-1$.
\end{corollary}

\section{Dequantization}
In this section we describe how to invert the Weyl quantization procedure on the torus.  Namely, given a linear operator $A$ acting on  $\C^N$, we want to associate to it a function $\alpha : \T^2 \to \C$ reproducing $A$ under Weyl quantization. By the non-injectivity of the Weyl quantization $\alpha$ will not be unique. Actually, we characterize all functions  admitting $A$ as Weyl operator.  
\begin{theorem}\label{th:deq}
Let $A$ be a linear operator in $\C^N$ represented by the matrix $(A_{n,j})_{n,j=0,\dots,N-1}$. Consider $\tilde{W}_{ N}(A)$,  defined in (\ref{Wignertens2}), and  the $(\theta,N)$-sampling operator $\mu_{\theta,N}$ defined in Definition (\ref{samplingoperator}). Then $A= \textrm{Op}^W_{\theta, N}(\alpha)$ for all $\alpha: \T^2 \to \C$ such that $\mu_{\theta,N}(\alpha)=N\tilde{W}_{ N}(A)$. 
\end{theorem}
\proof
Let $\alpha: \T^2 \to \C$ such that $\mu_{\theta,N}(\alpha)=N\tilde{W}_{ N}(A)$. We have to prove that $A_{n,j}=\langle u_n, \textrm{Op}^W_{\theta, N}(\alpha) u_j\rangle$, for all $n,j =0, \dots, N-1$. By (\ref{matrixcoeffWeylquant}) in Theorem~\ref{th:weylquant} we have that
\begin{eqnarray}\label{calc:dequant}
&& \langle u_n, \textrm{Op}_{\theta,N}^W(\alpha) u_j \rangle \nonumber \\
 &=& \frac{1}{2N}\left( \mathcal{F}_2\alpha_{\theta,N}(j+n,j-n)+ \mathcal{F}_2\alpha_{\theta,N}(j+n+N,j-n+N) \right)  \nonumber \\
&=& \frac{1}{2}\sum_{n_2 =0}^{2N-1}e^{-\frac{2\pi in_2(j-n)}{2N}}\left[\tilde{W}_N (A)(j+n,n_2)+  (-1)^{n_2}\tilde{W}_N(A)(j+n+N,n_2)\right]\nonumber   \\
&=&\sum_{n_2 =0}^{2N-1}e^{-\frac{2\pi in_2(j-n)}{2N}}\tilde{W}_N(A)(j+n,n_2)  \nonumber \\
&=&  \sum_{l=0}^{N-1}A_{l,j+n-l}\delta^{(2N)}_{2l,2n}=A_{n,j}. \nonumber
\end{eqnarray}
where we used the symmetry of the Wigner transform  (\ref{prop:Wignaone}). This concludes the proof of the Theorem.\qed
\begin{corollary}\label{alphaewigner}
Let $\alpha: \T^2 \to \C$   a function on $\T^2$,  and  $A=\textrm{Op}^W_{\theta, N}(\alpha)$ its Weyl operator in the representation $(\theta,N)$. Then for all $\alpha': \T^2 \to \C$ such that $\mu_{\theta,N}(\alpha')=N\tilde{W}_{ N}(A)$ it results that $\alpha' \doteq_{\theta,N} \alpha$.
\end{corollary}
\begin{remark}\label{remark:deq1}
Let us discuss in what sense Theorem~\ref{th:deq} represents an inversion of the Weyl of the quantization procedure. Consider a function $\alpha \in {\mathcal F}(\T^2)$ and its Weyl operator $A:=\textrm{Op}_{\theta,N}^W(\alpha)$. Dequantize $A$ using Theorem~\ref{th:deq} (which amounts to compute its Wigner transform, namely $\tilde{W}_{ N}(A)$). The question is whether or not $\mu_{\theta,N}(\alpha)$ is equal  to $N\tilde{W}_{ N}(A)$, namely whether the values assumed by $\alpha$ on $L(\theta,N)$ are equal or not to the values of $N\tilde{W}_{N}(A)$. The answer is negative simply because the coefficients $\{N\tilde{W}_{N}(A)(r,s)\}_{r,s \in \Z_{2N}}$ satisfy the symmetries (\ref{prop:Wignaone}), (\ref{prop:Wignatwo}), (\ref{prop:Wignathree}),  not satisfied in general by  $\{\alpha_{\theta,N}(r,s)\}_{r,s \in \Z_{2N}}$.   The next result, Corollary \ref{cor:muwignervalues},  entails a kind of universality for the sum of the values assumed on the lattice having $4$ points  at a distance $N$. More precisely: 
\end{remark}

\begin{corollary}\label{cor:muwignervalues}
Let  $\alpha, \alpha' :\T^2 \to \C $, then the following proposition are equivalent:
\begin{enumerate}
\item $\alpha \doteq_{\theta,N} \alpha'$;
\item for all $ r,s=0, \dots,N-1$:
$$
\Delta(\mu_{\theta,N}(\alpha))_{r,s}=\Delta(\mu_{\theta,N}(\alpha'))_{r,s}=4N \tilde{W}_N(A)(r,s),
$$
where $A=\textrm{Op}^W_{\theta, N}(\alpha)=\textrm{Op}^W_{\theta, N}(\alpha')$.
 \end{enumerate}
\end{corollary}
\proof
The proof can be obtained combining Theorem \ref{th:characteq} and Corollary \ref{alphaewigner}.\qed
\begin{remark}
By Corollary \ref{cor:muwignervalues}, it follows that the equivalence class of symbols related to the same Weyl operator $A$ is completely characterized by the principal sub-matrix of $\tilde{W}_N(A)$, (the $N \times N$ sub-matrix extract by the $2N \times 2N$ matrix $\tilde{W}_N(A)$ taking the first $N$ rows and columns), namely the values assumed on the independent lattice $I(\theta,N).$
\end{remark}
\begin{remark}\label{remarknotinjective2}
Let us sum up some relevant aspects about  Weyl quantization and dequantization on the torus:
\begin{itemize}
\item[(i)]
Each equivalence class in   ${\mathcal F}(\T^2)$ contains at least one element having the symmetries described in Corollary \ref{cor:symmetriesWF}  on the lattice $L(\theta,N) $. Namely, given  $\alpha\in{\mathcal F}(\T^2)$  there exists  $\alpha^\prime \doteq_{\theta,N} \alpha$  such that for all $m,l =0, \dots, N-1$
\begin{eqnarray}
\alpha^\prime_{\theta,N}\left(m+N,l\right)&=&(-1)^l\alpha^\prime_{\theta,N}\left(m,l\right), \nonumber \\ 
\alpha^\prime_{\theta,N}\left(m,l+N\right)&=&(-1)^m\alpha^\prime_{\theta,N}\left(m,l\right),  \nonumber \\
\alpha^\prime_{\theta,N}\left(m+N,l+N\right)&=&(-1)^{m+l+N}\alpha^\prime_{\theta,N}\left(m,l\right), \nonumber
\end{eqnarray}
where 
\begin{equation}
\alpha^\prime_{\theta,N}\left(r,s\right):=\alpha^\prime\left(\frac{r}{2N}+\frac{\theta_1}{ N}, \frac{s}{2N}+\frac{\theta_2}{ N}\right), \quad r,s =0,\dots,2N-1. \nonumber
\end{equation}
\item[(ii)] The same equivalence class contains functions with different values on $L(\theta,N)$.
\end{itemize}
\end{remark}

\section{Spin in phase space} 
In this section we consider in detail the case of a spin $1/2$, namely $N=2$. In particular we analyze the Wigner transform and its $4$ independent values and we show how they are related to the Pauli matrices.
\subsection{Wigner transform and Pauli matrices}\label{sub:Wig2}
Let $N=2$,   $\psi , \varphi \in \C^2$. By  (\ref{explicitWigner}), for all $\theta=(\theta_1,\theta_2) \in \T^2$ we have: 
\begin{equation}
W_{\theta,2}(\psi,\varphi)(x,p)=\sum_{m,n=0}^{3}\tilde{W}_2(\psi,\varphi)(m,n)\; \delta_{\Z} \left( x-\frac{m}{4}-\frac{\theta_1}{2}\right) \delta_{\Z} \left( p-\frac{n}{4}-\frac{\theta_2}{2}\right) , \nonumber
\end{equation}
where
\begin{equation}
\tilde{W}_2(\psi,\varphi)(m,n)=\frac{1}{4}\left[ \overline{\psi}_{m}\varphi_0e^{\frac{\pi im n}{2}} +\overline{\psi}_{m-1}\varphi_1e^{\frac{\pi i(m-2) n}{2}}  \right]. \nonumber
\end{equation}
Now construct the matrix $\mathbf{\tilde{W}_2 (\psi,\varphi)} :=(\tilde{W}_2(\psi,\varphi)(m,n))_{m,n=0,1,2,3}$ having as entries the values of $\tilde{W}_2(\psi,\varphi)$.  A simple computation yields:
\begin{equation}\label{Wigner2}
\mathbf{\tilde{W}_2 (\psi,\varphi)} = \frac{1}{4}\left( \begin{array}{cccc}  \langle \psi,I \varphi  \rangle &   \langle \psi, \sigma_z \varphi  \rangle &   \langle \psi,I   \varphi  \rangle &  \langle \psi, \sigma_z \varphi  \rangle  \\ 
 \langle \psi, \sigma_x\varphi  \rangle &   \langle \psi, \sigma_y \varphi  \rangle & -  \langle \psi, \sigma_x \varphi  \rangle & -  \langle \psi, \sigma_y\varphi  \rangle \\
  \langle\psi, I \varphi  \rangle & -  \langle \psi, \sigma_z \varphi  \rangle   &  \langle \psi,I \varphi  \rangle &  - \langle \psi, \sigma_z \varphi  \rangle  \\
 \langle \psi, \sigma_x \varphi  \rangle & -  \langle \psi, \sigma_y \varphi  \rangle  &-   \langle \psi,\sigma_x \varphi  \rangle &  \langle \psi,\sigma_y \varphi  \rangle
\end{array} \right)
\end{equation}
where $I$ is the identity matrix and $\sigma_x, \sigma_y, \sigma_z$ are the Pauli matrices, namely
$$ 
\textrm{I} = \left( \begin{array}{ccc} 1 & \;& 0 \\ 
0 &\; & 1 
\end{array} \right), \qquad \sigma_x = \left( \begin{array}{ccc} 0 &\; & 1 \\ 
1 & \;& 0 
\end{array} \right),
$$
$$
\sigma_y = \left( \begin{array}{ccc} 0 &\; & -i \\ 
i &\; & 0 
\end{array} \right), \qquad \sigma_z = \left( \begin{array}{ccc} 1 & \;& 0 \\ 
0 &\; & -1 
\end{array} \right).
$$
As explained, the independent values of $\tilde{W}_2 (\psi,\varphi)$ are $\{\tilde{W}_2 (\psi,\varphi)(m,n)\}_{m,n=0,1}$  and they correspond to the first $2\times 2 $ block in the matrix $\mathbf{\tilde{W}_2} (\psi,\varphi)$. We denote this sub-matrix $\mathbf{\tilde{w}_2}(\psi,\varphi)=(\tilde{W}_2 (\psi,\varphi)(m,n))_{m,n=0,1}$ and we call it the \emph{principal sub-matrix}. It results that
$$
\mathbf{\tilde{w}_2}(\psi,\varphi) = \frac{1}{2}\left( \begin{array}{cc}  \langle \psi,I \varphi \rangle  & \langle \psi, \sigma_z \varphi \rangle  \\ 
  \langle \psi, \sigma_x \varphi \rangle&  \langle \psi, \sigma_y\varphi \rangle
\end{array} \right).
$$
In the case $\varphi= \psi$ the matrix corresponding to $\mathbf{\tilde{W}_2 \psi} =(\tilde{W}_2\psi(m,n))_{m,n=0,1,2,3}$ is given by
$$
\mathbf{\tilde{W}_2 \psi} = \frac{1}{4}\left( \begin{array}{cccc}  \langle \psi,I \psi  \rangle &   \langle \psi, \sigma_z \psi  \rangle &   \langle \psi,I   \psi \rangle &  \langle \psi, \sigma_z \psi  \rangle  \\ 
 \langle \psi, \sigma_x \psi \rangle &   \langle \psi, \sigma_y \psi \rangle & -  \langle \psi, \sigma_x \psi  \rangle & -  \langle \psi, \sigma_y\psi \rangle \\
  \langle\psi, I \psi \rangle & -  \langle \psi, \sigma_z \psi \rangle   &  \langle \psi,I\psi  \rangle &  - \langle \psi, \sigma_z \psi \rangle  \\
 \langle \psi, \sigma_x \psi \rangle & -  \langle \psi, \sigma_y \psi  \rangle  &-   \langle \psi,\sigma_x \psi \rangle &  \langle \psi,\sigma_y \psi \rangle
\end{array} \right)
$$
and the principal sub-matrix is $\mathbf{\tilde{w}_2}\psi=(\tilde{W}_2 \psi(m,n))_{m,n=0,1}$
$$
\mathbf{\tilde{w}_2}\psi=\frac{1}{2} \left( \begin{array}{cc}  \langle \psi,I \psi  \rangle  &  \langle \psi, \sigma_z \psi  \rangle  \\ 
  \langle \psi, \sigma_x \psi \rangle&  \langle \psi, \sigma_y \psi \rangle
\end{array} \right).
$$
Therefore, the state $\psi$ is determined by the expectation values of the Pauli matrices on the vector $\psi$.  Here  the knowledge of the values of  $W_{\theta,2}\psi$ in the ghost lattice is critical,  because the values corresponding to the even indices in the matrix $\mathbf{\tilde{W}_2 \psi}$ are four times  $\|\psi\|^2$. Therefore the  relevant information about the state $\psi$ is encoded in the values of $W_{\theta,2}\psi$ on the ghost lattice.
\begin{remark}
In (\ref{Wigner2}) the support properties of the two marginals described in (\ref{supp1margdisc}) and (\ref{supp2margdisc}) of Proposition \ref{propTdisc} are apparent. The sum  of all  rows (first marginal) is actually $0$ in correspondence of columns  with odd indces  ($1$ and $3$). The same is true summing all the columns (second marginal): we obtain $0$ in correspondence of rows with odd indices. 
\end{remark}

\subsection{Dequantization of the Pauli matrices}\label{sub:deqpauli}
We want to compute the classical symbols of the Pauli matrices and of the identity matrix.

According to Theorem~\ref{th:deq} and Corollary~\ref{cor:muwignervalues}, we can say that the Weyl symbols  of the matrices $\{I,\sigma_x,\sigma_y,\sigma_z\}$ in the representation $(\theta,2)$ can be obtained simply computing the corresponding Wigner transform, namely $ \tilde{W}_2(I)$ and  $ \tilde{W}_2(\sigma_j)$, $j=x, y, z$. 
We compute explicitly the matrices
$$
\mathbf{\tilde{W}_2}(I)=(\tilde{W}_2(I)(m,n))_{m,n =0,1,2,3}
$$
and
$$
\mathbf{\tilde{W}_2}(\sigma_j)=(\tilde{W}_2(\sigma_j)(m,n))_{m,n =0,1,2,3} \quad j=x,y,z
$$
that completely characterize the (equivalence class of) symbols of the matrices $I,\sigma_x,\sigma_y, \sigma_z$. The result is:
$$
\mathbf{\tilde{W}_2}(I)=\frac{1}{2} \left( \begin{array}{ccccccc} 1 &\; & 0 &\; & 1 &\; & 0 \\ 
0 & \; &0 &\; & 0 &\; & 0 \\
1 &\; & 0 &\; & 1 &\; & 0 \\
0 & \; &0 &\; & 0 &\; & 0
\end{array} \right), \qquad \mathbf{\tilde{W}_2}(\sigma_x)= \frac{1}{2} \left( \begin{array}{ccccccc} 0 &\; & 0 &\; & 0 &\; & 0 \\ 
1 &\; & 0 &\; & -1 &\; & 0 \\
0 & \; &0 &\; & 0 &\; & 0 \\
1 &\; & 0 &\; & -1 &\; & 0
\end{array} \right),
$$
$$
\mathbf{\tilde{W}_2}(\sigma_y)=\frac{1}{2}  \left( \begin{array}{ccccccc} 0 &\; & 0 &\; & 0 &\; & 0 \\ 
0 & \; &1 &\; & 0 & \; &-1 \\
0 &\; & 0 &\; & 0 &\; & 0 \\
0 &\; & -1 &\; & 0 &\; & 1
\end{array} \right), \qquad \mathbf{\tilde{W}_2}(\sigma_z)=\frac{1}{2}  \left( \begin{array}{ccccccc} 0 &\; & 1 &\; & 0 &\; & 1 \\ 
0 &\; & 0 &\; & 0 &\; & 0 \\
0 &\; & -1 &\; & 0 &\; & -1 \\
0 &\; & 0 &\; & 0 &\; & 0
\end{array} \right),
$$
so we have that their principal sub-matrix correspond, up a multiplication factor $\frac{1}{2}$, to the canonical basis in the space $M_2(\C)$:  
$$
\mathbf{\tilde{w}_2}(I)=\frac{1}{2} \left( \begin{array}{ccc} 1 & \; & 0  \\ 
0 &\; & 0  
\end{array} \right), \qquad \mathbf{\tilde{w}_2}(\sigma_x)= \frac{1}{2} \left( \begin{array}{ccc} 0 &\; & 0 \\ 
1 & \; & 0 
\end{array} \right),
$$
$$
\mathbf{\tilde{w}_2}(\sigma_y)=\frac{1}{2}  \left( \begin{array}{ccc} 0 &\; & 0  \\ 
0 &\; & 1 
\end{array} \right), \qquad \mathbf{\tilde{w}_2}(\sigma_z)=\frac{1}{2}  \left( \begin{array}{ccc} 0 &\; & 1 \\ 
0 &\; & 0
\end{array} \right).
$$
\begin{theorem}\label{Th:charaymb2pauli}
The Weyl symbols of $I, \sigma_x, \sigma_y \sigma_z$ can be characterized as follows:
\begin{enumerate}
\item $\alpha_I: \T^2 \to \C$ is (up to equivalence) the Weyl symbol in the representation $(\theta,2)$ of $I$ if and only if 
$$
\Delta(\mu_{\theta,2}(\alpha_I))=4 \left( \begin{array}{ccc} 1 & \; & 0  \\ 
0 &\; & 0  
\end{array} \right);
$$
\item $\alpha_x: \T^2 \to \C$ is (up to equivalence) the Weyl symbol in the representation $(\theta,2)$ of $\sigma_x$ if and only if 
$$
\Delta(\mu_{\theta,2}(\alpha_x))=4 \left( \begin{array}{ccc} 0 & \; & 0  \\ 
1 &\; & 0  
\end{array} \right);
$$
\item $\alpha_y: \T^2 \to \C$ is (up to equivalence) the Weyl symbol in the representation $(\theta,2)$ of $\sigma_y$ if and only if 
$$
\Delta(\mu_{\theta,2}(\alpha_y))=4 \left( \begin{array}{ccc} 0 & \; & 0  \\ 
0 &\; & 1  
\end{array} \right);
$$
\item $\alpha_z: \T^2 \to \C$ is (up to equivalence) the Weyl symbol in the representation $(\theta,2)$ of $\sigma_z$ if and only if 
$$
\Delta(\mu_{\theta,2}(\alpha_z))=4 \left( \begin{array}{ccc} 0 & \; & 1  \\ 
0 &\; & 1  
\end{array} \right).
$$
\end{enumerate}
\end{theorem}
\proof The proof is a direct consequence of Theorem~\ref{th:deq} and Corollary \ref{cor:muwignervalues}. \qed
\begin{remark}
It is well known that the Pauli matrices are related to the generators of the $2$-dimensional Weyl system, see \cite{Schwinger}. More precisely it is easy to see that  
\begin{equation}\label{sigmaxz}
\sigma_z=e^{-\pi i \theta_1}T_{\theta,2}(1,0) \quad \textrm{and} \quad  \sigma_x=e^{-\pi i \theta_2}T_{\theta,2}(0,1) 
\end{equation}
and that 
\begin{equation}\label{sigmay}
\sigma_y=e^{-\pi i (\theta_1+\theta_2)}T_{\theta,2}(1,1).  
\end{equation}
Using this relations it is easy to compute (up to equivalence) the Weyl symbols $\alpha_I,\alpha_x, \alpha_y, \alpha_z$ of $I, \sigma_x, \sigma_y, \sigma_z$ respectively without using the dequantization procedure, i.e. Theorem \ref{th:deq}:
\end{remark}
\begin{theorem}
Let $\alpha_I, \alpha_x, \alpha_y, \alpha_z: \T^2 \to \C$ such that for all $(x,p) \in \T^2$:
\begin{equation}
\alpha_I(x,p)=1, \quad \alpha_x(x,p)=e^{2\pi i \left( p- \frac{\theta_2}{2}\right)}
\end{equation}
and
\begin{equation}
\alpha_y(x,p)=e^{2\pi i \left( x- \frac{\theta_1}{2}\right)} e^{2\pi i \left( p- \frac{\theta_2}{2}\right)}, \quad
\alpha_z(x,p)=e^{2\pi i \left( x- \frac{\theta_1}{2}\right)}.
\end{equation}
Then $\textrm{Op}^W_{\theta,N}(\alpha_I)=I$, $\textrm{Op}^W_{\theta,N}(\alpha_j)=\sigma_j$, $j=x,y,z$.
\end{theorem}
\proof
The proof is an immediate consequence of (\ref{sigmaxz}, \ref{sigmay}) and (\ref{symbolsof generators}), or it can be obtained using Theorem \ref{Th:charaymb2pauli}.\qed
\begin{remark}
According to Theorem \ref{Th:charaymb2pauli}, the Weyl quantization produces a natural correspondence between the canonical basis of $M_2(\C)$ and the matrices $\{I, \sigma_x, \sigma_y, \sigma_z\}$. This correspondence suggests how to define the natural extension of the Pauli matrices in dimension $N >2$.
\end{remark}
\subsection{$N$-dimensional Pauli matrices}\label{Npauli}
We want to generalize the contents of subsections \ref{sub:Wig2}, \ref{sub:deqpauli} to   $N>2$. Let $\theta=(\theta_1, \theta_2) \in \T^2$ and $\psi, \varphi \in \C^N$,  from (\ref{explicitWigner}) it follows, by a simple computation, that for all $r,s=0, \dots, 2N-1$: 
\begin{equation}
\tilde{W}_{N}(\psi,\varphi)(r,s)=\frac{1}{2N}\langle \psi, B^{[r,s]} \varphi \rangle, \nonumber
\end{equation}
where 
\begin{equation}\label{Arsdef}
B^{[r,s]}=\sum_{j = 0}^{N-1}e^{-\frac{\pi i(r-2j)s}{N}}E_{j,r-j}^{(N)}, \nonumber 
\end{equation}
and $(E^{(N)}_{k,n})_{m,l}=\delta_{k,m}^{(N)}\delta_{n,l}^{(N)}$ is the canonical basis of $M_N(\C)$. Define the matrix $\mathbf{\tilde{W}}_N(\psi,\varphi)=(\tilde{W}_{N}(\psi,\varphi)(r,s))_{r,s=0,\dots, 2N-1}$ in the following way:
\begin{equation}
\mathbf{\tilde{W}}_N(\psi,\varphi)=\left(\frac{1}{2N}\langle \psi, B^{[r,s]} \varphi \rangle \right)_{r,s=0,\dots, 2N-1}.
\end{equation}
According to (\ref{propWigner1}), (\ref{propWigner2}), (\ref{propWigner3}), the matrices $(B^{[r,s]})_{r,s=0,\dots, 2N-1}$ have the following properties: for all $r,s =0,\dots,2N-1$
 \begin{equation}
B^{[r+N,s]}=(-1)^sB^{[r,s]}, \; \; \; B^{[r,s+N]}=(-1)^rB^{[r,s]}, 
\end{equation}
and 
\begin{equation}
B^{[r+N,s+N]}=(-1)^{r+s+N}B^{[r,s]},
\end{equation}
where the sums in the upperscript are modulo $2N$.  Thus the independent matrices are those labelled with $r,s=0,\dots,N-1$. These $N^2$ matrices are a natural generalization to higher dimensional spaces of the Pauli matrices because, as in the $2$-dimensional case, they correspond, via  Weyl quantization,  to the equivalence class of symbols specified by the canonical basis of $M_{N}(\C)$. More precisely:
\begin{theorem}\label{th:weyl symbols Npauli}
Let $\theta \in \T^2$ and $r,s=0,\dots,N-1$, then $\beta^{[r,s]}: \T^2 \to \C$ is (up to equivalence) the Weyl symbol in the representation $(\theta,N)$ of $B^{[r,s]}$ if and only if
\begin{equation}
\Delta(\mu_{\theta,N}(\beta^{[r,s]}))=2N E_{r,s}^{(N)}.
 \end{equation} 
\end{theorem}
\proof The proof is a direct consequence of Theorem~\ref{th:deq} and Corollary \ref{cor:muwignervalues}.\qed

We conclude this section with the explicit computation of $\beta^{[r,s]}$ (Weyl symbols of the $N$-dimensional Pauli matrices $B^{[r,s]}$, up to equivalence).
\begin{theorem}
Let  $r,s=0,\dots,N-1$ and $\beta^{[r,s]}: \T^2 \to \C$ such that for all $(x,p) \in \T^2$:
\begin{equation}
\beta^{[r,s]}(x,p)= \frac{1}{2N}\sum_{k,m=0}^{2N-1}e^{2\pi i k \left( x- \frac{r}{2N}-\frac{\theta_1}{N}\right)}e^{2\pi i m \left( p- \frac{s}{2N}-\frac{\theta_2}{N}\right)}.
\end{equation}
Then $\textrm{Op}^W_{\theta,N}(\beta^{[r,s]})=B^{[r,s]}$.
\end{theorem}
\proof
First we observe that 
$$
\mu_{\theta,N}(\beta^{[r,s]})_{j,l}=\beta^{[r,s]}\left(\frac{j}{2N}+\frac{\theta_1}{N},\frac{l}{2N}+\frac{\theta_2}{N}\right)=2N \delta^{(2N)}_{j,r}\delta^{(2N)}_{l,s},
$$
for all $j,l=0, \dots 2N-1$. Then it follows immediately that 
$$
\Delta(\mu_{\theta,N}(\beta^{[r,s]}))=2N E^{(N)}_{r,s}
$$ 
and so we can apply Theorem \ref{th:weyl symbols Npauli}.\qed

\section{Appendix}
In this appendix we recall some basic results on the discrete Heisenberg group $\mathbb{H}(\Z) $ and its unitary irreducible representations. There are different (equivalent) approaches to the classifications of the finite dimensional representations of $\mathbb{H}(\Z)$, e.g \cite{de,Bou}, and here we refer to \cite{de}. 
The discrete Heisenberg group $\mathbb{H}(\Z)$ is given  by $\Z^2 \times \R$ with the following product
\begin{equation}
(n_1,n_2,s)(m_1,m_2,r)=\left(n_1+m_1, n_2+m_2, s+r-\frac{1}{2}(n_1m_2-n_2m_1)\right), \nonumber
\end{equation}
for all $(n_1,n_2,s),(m_1,m_2,r) \in \mathbb{H}(\Z)$. 
The problem of the classification of all the finite dimensional unitary and irreducible representations of $\mathbb{H}(\Z)$ has been addressed and solved, \cite{de,andpas}, and can be formulated as follows.
 
\begin{theorem}\label{theorem repr N}
Let $N \in \N$, $N > 1$, and let $\{u_{0}, u_1, \dots, u_{N-1}\}$ be the canonical basis of $\C^N$.
\begin{enumerate}
\item For all $\theta=(\theta_1,\theta_2)\in \T^2$,  the map $T_{\theta, N}:\mathbb{H}(\Z) \to \mathcal{U}(\C^N)$ such that for all $(n_1,n_2,s)  \in \mathbb{H}(\Z)$ and for all $j=0,\dots, N-1$
\begin{equation}\label{TNk}
T_{\theta, N}(n_1,n_2,s)u_j=e^{\frac{2\pi i s}{N}}e^{-\frac{\pi i n_1n_2}{N}}e^{\frac{2\pi i n_1\left(j+ \theta_1 \right)}{N}}e^{\frac{2 \pi i\theta_2 n_2}{N}}u_{j-n_2}
\end{equation}
where the result of $j-n_2$ is modulo  $N$, is a unitary irreducible representation of the discrete Heisenberg group $\mathbb{H}(\Z)$ on $\C^N$
\item Given $\theta, \tilde{\theta}\in \T^2$, the representations $T_{\theta, N}$ and $T_{\tilde{\theta}, N}$ are equivalent if and only if $\theta=\tilde{\theta}$.
\item If $\rho$ is any  unitary irreducible representation of $\mathbb{H}(\Z)$ with $\rho(0,0,s)=e^{\frac{2\pi i s}{N}}I$, then there exist a unique element $\theta\in \T^2$ such that $\rho$ is unitarily equivalent to $T_{\theta, N}$.
\end{enumerate}
\end{theorem}
We present some properties of the family of irreducible $N$-dimensional  representations $\{T_{\theta,N}\}_{\theta \in \T^2}$ defined in (\ref{TNk}), \cite{de}. First observe that, since the last variable $s$ always acts in a simple way as multiplication by the scalar $e^{\frac{2\pi i s}{N}}$, it is convenient to disregard it entirely, so for all $\theta=(\theta_1,\theta_2) \in \T^2$ we define, with an abuse of notation, the reduced map $T_{\theta,N}: \Z^2 \to \mathcal{U}(\C^N)$ such that for all $n_1,n_2 \in \Z$
\begin{equation}\label{def:T}
T_{\theta,N}(n_1,n_2)=T_{\theta,N}(n_1,n_2,0)=e^{-\frac{\pi i n_1n_2}{N}}t_2(\theta_2)^{n_2}t_1(\theta_1)^{n_1}, 
\end{equation}
where $t_1(\theta_1):=T_{\theta,N}(1,0,0)$ and $t_2(\theta_2):=T_{\theta,N}(0,1,0)$. 
\begin{proposition} \label{prop repr T}
Let   $\theta=(\theta_1,\theta_2) \in \T^2$, then:
\begin{enumerate}
\item for all $n_1,n_2\in \Z$
\begin{equation}
T_{\theta,N}(n_1,n_2)^*=T_{\theta,N}(-n_1,-n_2); \nonumber
\end{equation}
\item for all $n_1,n_2, m_1,m_2 \in \Z$
\begin{equation}
T_{\theta,N}(n_1,n_2)T_{\theta,N}(m_1,m_2)=e^{-\frac{\pi i (n_1m_2-n_2m_1)}{N}}T_{\theta,N}(n_1+m_1,n_2+m_2);\nonumber
\end{equation}
\item for all $n_1,n_2\in \Z$
\begin{equation}
t_1(\theta_1)^{n_1} t_2(\theta_2)^{n_2}=e^{-\frac{2\pi i n_1n_2}{N}}t_2(\theta_2)^{n_2} t_1(\theta_1)^{n_1};\nonumber
\end{equation}
\item  for all $n_1,n_2,j,l\in \Z$
\begin{equation}
T_{\theta,N}(n_1+2Nj_1,n_2+2Nj_2)=e^{2\pi i(2j_1\theta_1+2j_2 \theta_2)}  T_{\theta,N}(n_1,n_2);\nonumber
\end{equation}
\item for all $n,j \in \Z$
\begin{equation}
T_{\theta,N}(n+jN,0)= e^{2 \pi ij\theta_1} T_{\theta,N}(n,0)\nonumber
\end{equation}
and
\begin{equation}
 T_{\theta,N}(0,n+jN)= e^{ 2 \pi ij\theta_2} T_{\theta,N}(0,n).\nonumber
\end{equation}
\end{enumerate}
\end{proposition}
\begin{remark}
We observe that by assertion 5 of Proposition \ref{prop repr T} it follows that the operators $t_1(\theta_1)$ and  $t_2(\theta_2)$ are $N$-periodic, up to a phase factor, while, by assertion 4, that   $T_{\theta,N}$ is $2N$-periodic, up to a phase factor. This periodicity of $T_{\theta,N}$ is inherited by the Wigner transform and implies the double lattice support described in Remark \ref{remarksuppW}.
\end{remark}

\medskip

{\bf Acknowledgement.}
This work was supported by Fondazione Cassa di Risparmio di Puglia and by the National Group of Mathematical Physics (GNFM - INdAM).
\medskip

%and subsection a unique label (see Sect.~\ref{sec:1}).

\end{document}